\documentstyle[twocolumn,epsf,aps,eqsecnum]{revtex}         
\begin{document}
\widetext                                        
\title{Scattering in one dimension: The coupled Schr\"{o}dinger
equation, threshold behaviour and Levinson's theorem}
\author{K.A. Kiers}
\address{ Department of Physics, University of British
Columbia, \\ Vancouver, British Columbia, Canada \ V6T 1Z1}
\author{W. van Dijk}
\address{Redeemer College, Ancaster, Ontario, Canada L9G 3N6  \\
and Department of Physics and Astronomy, McMaster University,  \\
Hamilton, Ontario, Canada \ L8S 4M1}
\date{July 16, 1996}
\maketitle

\begin{abstract}
We formulate scattering in one dimension due to the coupled
Schr\"{o}dinger equation in terms of the
$S$ matrix, the unitarity of which leads to constraints on the scattering
amplitudes.  Levinson's theorem is seen to have the form $\eta(0) =
\pi (n_b + {1\over 2} n - {1\over 2} N)$, where $\eta(0)$ is the phase of
the $S$ matrix at zero energy, $n_b$ the number of bound states with
nonzero binding energy, $n$ the number of half-bound states,
and $N$ the number of coupled equations.  In view of the effects due to the
half-bound states, the threshold behaviour of the scattering
amplitudes is investigated in general, and is also illustrated by
means of particular potential models.
\end{abstract}


\section{Introduction}
\label{intro}
The quantum mechanics of one-dimensional scattering describes many
actual physical phenomena to a good approximation. (For example, see
Ref.~\cite{Hauge89} for a review of tunneling times.)  One-dimensional
models are furthermore often employed to gain deeper insight into the
approximations used in order to make the more complex
three-dimensional systems tractable.  It is therefore not surprising
that there have been many articles, also in this journal, dealing with
various aspects of such scattering.  In particular a number of papers
have appeared in recent years on the threshold behaviour of
one-dimensional scattering and Levinson's theorem
\cite{DeBianchi94,DeBianchi95,Senn88,Barton85,VanDijk92,Nogami95}.
These studies have been limited to systems without coupling.

In this paper we wish to investigate scattering described by a system
of coupled differential equations with a particular interest in
developing a formulation for Levinson's theorem and in gaining insight
into the threshold properties of the scattering amplitudes.  This work
can be seen as a special case of multichannel scattering for which the
threshold energies are equal.  In subsequent work, we
intend to generalize to the case of differing threshold energies.
Although in previous work on one-dimensional scattering one has at
times employed a ``partial wave'' analysis\cite{Nogami95,Eberly65} or
a parity-eigenstate representation\cite{DeBianchi94,VanDijk92}, we
have chosen to use the traditional, more ``physical'', approach
involving states with incident waves coming from a single direction.

In Sec.~\ref{smatrix} we express the scattering properties in terms of
the $S$ matrix, the unitarity of which leads to specifiable
constraints on the scattering amplitudes.  For the proof of the
generalized Levinson's theorem we make use of the complete set of
orthonormal states of the Hamiltonian; this is an alternative to the
approach involving the analyticity of the scattering
amplitudes.  The proof of Levinson's theorem depends on the threshold
properties of the scattering amplitudes.  These properties are of
interest in their own right, especially in connection with scattering
time delay and advance\cite{VanDijk92}, and therefore we discuss the
zero-energy behaviour of the amplitudes at some length.

The factorization of the $S$ matrix is generalized to the coupled system in
Sec.~\ref{segment}.  We also indicate that there is a class of
finitely periodic matrix potentials for which the scattering
amplitudes can be found in a way analogous to the case with no coupling.

In Sec.~\ref{model} we discuss a number of specific potential
models to elucidate and amplify general results.  We conclude
with a brief discussion of our results in Sec.~\ref{sec:disc}.

\section{$S$-matrix formulation}
\label{smatrix}
The one-dimensional scattering problem has been studied in terms of
the $S$ matrix by a number of authors.  (See, for example,
Refs.~\protect\cite{Nogami95,Faddeev64,Newton80,Newton83,Newton84}.)
We extend the formalism to include a matrix potential function.
The Schr\"{o}dinger equation for a stationary state of
such a system is
\begin{equation}
-\frac{d^2\Psi}{dx^2} + V(x) \Psi = k^2 \Psi,  \label{1}
\end{equation}
where $V(x)$ is a real, symmetric $N\times N$ matrix, $k^2$ the energy
of the system, and $\Psi(k,x)$ the wave function,
which is an $N$-dimensional
column vector.  For large values of $|x|$ the potential matrix $V(x)$
approaches zero sufficiently fast, so that in the asymptotic region
$\Psi(k,x)$ represents a free-particle wave function.  To ensure this
we will take $V(x) = {\bf 0}$ for $|x|>R$, $R$ being the range of the 
potential\footnote{The boldface {\bf
0} and {\bf 1} refer to the zero and identity $N\times N$ matrices,
respectively.}.  Furthermore, we assume that $|V_{ij}(x)|$ is
integrable for  $i,j=1,\dots,N$.

The physical scattering solutions of Eq.~(\ref{1}) at a given energy
$k^2$ can be written as the columns of $N\times N$ matrices
$\psi(k,x)$ and $\widetilde{\psi}(k,x)$ which are uniquely determined
by the boundary conditions,
\begin{equation}
\psi(k,x) \sim \cases{ {\bf 1}e^{ikx} + \rho(k) e^{-ikx}, & 
                                 $ x \rightarrow -\infty$\cr
               \tau(k) e^{ikx}, & $x \rightarrow \infty$\cr}
 \label{2}
\end{equation}
and
\begin{equation}
\widetilde{\psi}(k,x) \sim \cases{\widetilde{\tau}(k) e^{-ikx},
& $x \rightarrow -\infty$\cr
{\bf 1}e^{-ikx} + \widetilde{\rho}(k) e^{ikx},& $x \rightarrow \infty$.\cr}
 \label{3}
\end{equation}
We will refer to $\psi$ and $\widetilde{\psi}$ as the solution matrices.
Note that the columns of $\psi$ contain the wave functions with an
incident wave from the left, whereas the columns of $\widetilde{\psi}$
includes those with an incident wave from the right.  The $N\times N$
matrices $\rho, \widetilde{\rho}, \tau, \widetilde{\tau}$ are
generalizations of the usual reflection and transmission
amplitudes\cite{VanDijk79,VanDijk79a}.  The set of $N$-dimensional
column vectors of matrices $\psi$ and $\widetilde{\psi}$ represent
solutions of Eq.~(\ref{1}).  The linear independency of these
solutions can be shown by considering the $2N\times 2N$ matrix,
\begin{equation}
W(\psi,\widetilde{\psi}) = \left( \begin{array}{cc} 
~\psi~ & ~\widetilde{\psi} \\  
\psi' & \widetilde{\psi}' \end{array} \right),   \label{3a}
\end{equation}
in which the prime indicates the derivative with respect to $x$.  In
Appendix A we show that the $\det{W(\psi,\widetilde{\psi})}$ is a
constant, which is nonzero if and only if the solutions are linearly
independent.  Its value, determined from the asymptotic forms of
$\psi$ and $\widetilde{\psi}$, is
\begin{equation}
\det{W(\psi,\widetilde{\psi})} = \det({\tau(k)})(-2ik)^N.   \label{3b}
\end{equation}
Thus when both $k$ and $\det{\tau(k)}$ are nonzero, the 
columns of $\psi$ and $\widetilde{\psi}$ give $2N$
linearly independent solutions.

In order to define the $S$ matrix, we consider the general
solution matrix in the asymptotic region,
\begin{equation}
\Phi(k,x) \sim \cases{ Ae^{ikx} + B'e^{-ikx}, & $x\rightarrow -\infty$\cr
                      A'e^{ikx} + Be^{-ikx}, & $x\rightarrow
\infty$,\cr}
\label{4}
\end{equation}
where $A,A',B,B'$ are $N\times N$ matrices.  Since the unprimed
matrices are associated with incoming waves and the primed matrices
with outgoing waves, the $S$ matrix can be defined as the matrix that
transforms the coefficients of the incoming waves into those of the
outgoing waves~\cite{Nogami95,Lane58}, so that   
\begin{equation}
\left( \begin{array}{c}
       A' \\ B'
       \end{array} 
\right) = S \left( \begin{array}{c}
       A  \\ B
       \end{array}
\right) = \left( \begin{array}{cc}
       S_{aa} & S_{ab}  \\  S_{ba} & S_{bb} 
                 \end{array}
\right) \left( \begin{array}{c} A \\ B \end{array} \right). \label{5}
\end{equation}
Clearly, $S$ is a $2N \times 2N$ matrix. We can write it in terms of
the transmission and reflection amplitudes by making use of the
special cases for which $(A,B) = (\bf{1},\bf{0})$ and $(A,B) =
(\bf{0},\bf{1})$.  The result is
\begin{equation}
S = \left( \begin{array}{cc} 
S_{aa} & S_{ab} \\ S_{ba} & S_{bb}
\end{array} 
\right) 
= \left( \begin{array}{cc} ~\tau~ & ~\widetilde{\rho}~ \\
~\rho~ & ~\widetilde{\tau}~ \end{array} \right).  \label{6}
\end{equation}  
The $S$ matrix contains $4N^2$ complex elements or $8N^2$ real
parameters.  As we will see below there are a number of relations
between the transmission and reflection amplitudes, which will reduce
the number of independent real parameters.  

\subsection{Relation  between reflection and transmission amplitudes}

Constraints on the transmission and reflection amplitudes
follow from the Schr\"{o}dinger equation~(\ref{1}).  Consider two
solution matrices $\psi_1(k,x)$ and $\psi_2(k,x)$, then
\begin{equation}
\psi_2'' = (k^2 - V) \psi_2 \hspace{.2in} \mbox{and}
\hspace{.2in} \psi_2^{\dagger}{''} = \psi_2^\dagger (k^2 - V), \label{7}
\end{equation}
so that
\begin{equation}
\psi_2^{\dagger}{''} \psi_1 = \psi_2^\dagger (k^2 - V)
\psi_1. \label{8}
\end{equation}
The Schr\"{o}dinger equation, Eq. (\ref{1}), for $\psi_1$
premultiplied by $\psi_2^\dagger$ yields
\begin{equation}
\psi_2^\dagger \psi_1'' = \psi_2^\dagger (k^2 - V)
\psi_1. \label{9}
\end{equation}
Subtracting Eqs. (\ref{8}) and (\ref{9}), we obtain
\begin{equation} 
\psi_2^\dagger{''} \psi_1 - \psi_2^\dagger \psi_1'' =
\frac{d~}{dx} \left[ \psi_2^\dagger{'} \psi_1 - \psi_2^\dagger \psi_1'
\right] = 0,  \label{10}
\end{equation}
which leads to
\begin{equation}
\psi_2^\dagger{'} \psi_1 - \psi_2^\dagger \psi_1' = \mbox{constant matrix.} 
  \label{11}
\end{equation}
If we now insert the asymptotic forms of $\psi$ or $\widetilde{\psi}$ for 
$\psi_1$ or $\psi_2$ into Eq.~(\ref{11}) and equate the
expression at $-\infty$ to that at $+\infty$, we obtain the following
relations.  
\begin{eqnarray}
\tau^\dagger \tau + \rho^\dagger \rho & =  
\widetilde{\tau}^\dagger \widetilde{\tau} + \widetilde{\rho}^\dagger
\widetilde{\rho} & =  \bf{1}, \label{13}  \\
\rho^\dagger \widetilde{\tau} + \tau^\dagger \widetilde{\rho} & =
\widetilde{\tau}^\dagger \rho + \widetilde{\rho}^\dagger \tau & =  \bf{0}.
\label{15} 
\end{eqnarray}
Eqs.~(\ref{13}) and (\ref{15}) are equivalent to the statement that
$S^\dagger S = I$, where $I$ is the $2N\times 2N$ identity matrix.

Further relations are found by using the time-reversal symmetry of the
system.  Since $V(x)$ is real, the complex conjugate solution
matrices $\psi^*$ and $\widetilde{\psi}^*$ are also solutions of the
Schr\"{o}dinger equation.  By complex conjugating Eq.~(\ref{4}), we
see that the roles of the incoming and outgoing asymptotic waves are
reversed, and indeed Eq.~(\ref{5}) is valid when $A\rightarrow B'^*$,
$B\rightarrow A'^*$, $A'\rightarrow B^*$, and $B'\rightarrow A^*$.
Thus Eq.~(\ref{5}) may be written as 
\begin{equation}
\left( \begin{array}{c} B^* \\ A^* \end{array} \right) = S \left(
\begin{array}{c} B'^* \\ A'^* \end{array} \right).  \label{16}
\end{equation}
Multiplying on the left by $S^\dagger$ and using $S^\dagger S = I$,
we find that
\begin{equation}
\left( \begin{array}{c} B'^* \\ A'^* \end{array} \right) = S^\dagger
\left( \begin{array}{c} B^*  \\ A^*  \end{array} \right),  \label{17}
\end{equation}
which leads to
\begin{equation}
\left( \begin{array}{c} A' \\ B' \end{array} \right)
 = \left( \begin{array}{cc} ~\bf{0}~ & ~\bf{1}~ \\ ~\bf{1}~ & ~\bf{0}~ \end{array} \right)
S^T \left( \begin{array}{cc} ~\bf{0}~ & ~\bf{1}~ \\ ~\bf{1}~ & ~\bf{0}~ \end{array} \right)
\left( \begin{array}{c} A \\ B \end{array} \right),  \label{18}
\end{equation}
where $S^T$ is the transpose of $S$.
Thus
\begin{equation} 
S = \left( \begin{array}{cc} ~\bf{0}~ & ~\bf{1}~ \\ ~\bf{1}~ & ~\bf{0}~ \end{array} \right)
S^T \left( \begin{array}{cc} ~\bf{0}~ & ~\bf{1}~ \\ ~\bf{1}~ &  ~\bf{0}~ \end{array} \right),
\label{19}
\end{equation}
from which it follows that
\begin{equation}
\widetilde{\tau} = \tau^T, \hspace{.2in} \rho = \rho^T, \hspace{.2in}
\mbox{and} \hspace{.2in} \widetilde{\rho}=\widetilde{\rho}^T. \label{20}
\end{equation}
Inserting these expressions into Eqs.~(\ref{13}) and (\ref{15}),
we obtain
\begin{eqnarray}
\widetilde{\tau} \widetilde{\tau}^\dagger + \rho \rho^\dagger & = 
\tau \tau^\dagger + \widetilde{\rho}\widetilde{\rho}^\dagger & = \bf{1},
\label{22} \\
\tau \rho^\dagger + \widetilde{\rho} \widetilde{\tau}^\dagger & =
\rho \tau^\dagger + \widetilde{\tau} \widetilde{\rho}^\dagger  & =
\bf{0}. \label{24} 
\end{eqnarray}
These equations yield the other half of the unitarity condition of the
$S$ matrix, so that
\begin{equation}
S^\dagger S = S S^\dagger = I.  \label{25}
\end{equation}
For a parity-invariant potential function, i.e., $V(-x) = V(x)$, there are
further constraints on the transmission and reflection amplitudes.  In
that case the amplitudes are symmetric matrices and the two types of 
amplitudes are the same, i.e.,
\begin{equation}
\rho = \widetilde{\rho} = \rho^T = \widetilde{\rho}^T, \hspace{.2in}
\mbox{and} \hspace{.2in}
\tau = \widetilde{\tau} = \widetilde{\tau}^T = \tau^T. \label{27}
\end{equation}

There is also a useful relation between the scattering amplitudes at
$k$ and at $-k$, which is easily obtained by generalizing the 
result for uncoupled potentials~\cite{Nogami95}.  Since $\psi^*(-k,x)$ and
$\widetilde{\psi}^*(-k,x)$ are solution matrices of the Schr\"{o}dinger
equation with the same boundary conditions as $\psi(k,x)$ and
$\widetilde{\psi}(k,x)$ respectively, we find that
\begin{eqnarray}
\tau^*(-k) & =  \tau(k), \; \; \rho^*(-k) & =\rho(k), \nonumber \\
\widetilde{\tau}^{\ *}(-k) & = \widetilde{\tau}(k), \; \;
\widetilde{\rho}^{\ *}(-k) & = \widetilde{\rho}(k). \label{27a}
\end{eqnarray} 
From these relations it follows immediately that the reflection and
transmission amplitudes at threshold (i.e., $k=0$) are real.

\subsection{Levinson's Theorem}

Levinson's theorem in its most common formulation for a spherically
 symmetric potential gives a relationship of the scattering phase
 shifts at zero and infinite energy.  The theorem has also been
 studied for one-dimensional systems without
 coupling~\cite{DeBianchi94,Barton85}).  We generalize the theorem to
 the matrix-potential case.  Levinson's theorem is a consequence of
 the orthogonality and completeness relation of the eigenstates of the
 total Hamiltonian.

The scattering states of the Schr\"{o}dinger equation (\ref{1})
defined by Eqs.~(\ref{2}) and (\ref{3}) along with the bound states
can be used to form a complete orthonormal set of eigenstates.
Suppose that there are $n_b$ bound states whose orthonormal wave
functions are denoted by the column vectors $\psi_{E_j}(x)$ with
$E_j(<0)$ referring to the bound state energy.  In case of degenerate
bound states, we label the state vectors with subscript $E_j$, but
allow the possibility of different subscripts $j$ for the same energy
in order to include all independent bound-state vectors.  For example,
one could have $E_j=E_i$ where $i\neq j$ when $E_i$ is a degenerate
energy eigenvalue.  The orthonormality
relations are
\begin{eqnarray}
\frac{1}{2\pi}&&\int_{-\infty}^\infty dx \;
\psi_j^\dagger(k,x)\psi_i(k',x)   \nonumber \\
&&= \frac{1}{2\pi}\int_{-\infty}^\infty dx \; \widetilde{\psi}_j^\dagger(k,x)
\widetilde{\psi}_i(k',x)  =   \delta_{ij} \delta(k'-k), \label{29}
\end{eqnarray}
\begin{eqnarray}
\frac{1}{\sqrt{2\pi}}&&\int_{-\infty}^\infty dx
\;\psi_j^\dagger(k,x)\psi_{E_i}(x) \nonumber \\
&& =\frac{1}{\sqrt{2\pi}}\int_{-\infty}^\infty dx \;
\widetilde{\psi}_j^\dagger(k,x)\psi_{E_i}(x) =
0, \label{32}
\end{eqnarray}
\begin{eqnarray}
&&\frac{1}{2\pi}\int_{-\infty}^\infty dx \;
\widetilde{\psi}_j^\dagger(k,x)\psi_i(k',x)  = 0 \; \mbox{~and~}
\nonumber \\
&&\int_{-\infty}^\infty dx \; \psi_{E_j}^\dagger(x)\psi_{E_i}(x)  =
\delta_{ij}, \label{33}
\end{eqnarray} 
where $\psi_i$ and $\widetilde{\psi}_i$ are the $i^{\rm th}$ columns of the
$\psi$ and $\widetilde{\psi}$ matrices respectively.  Thus the 
completeness relation is
\begin{eqnarray}
&& \sum_{j=1}^{n_b} \psi_{E_j}(x) \psi_{E_j}^\dagger(x') + \sum_{i=1}^N
\frac{1}{2\pi} \int_0^\infty dk \; \psi_i(k,x)
\psi_i^\dagger(k,x') \nonumber \\ 
&& + \sum_{i=1}^N \frac{1}{2\pi}
\int_0^\infty dk \; \widetilde{\psi}_i(k,x)
\widetilde{\psi}_i^\dagger(k,x') = {\bf 1} \delta(x-x') . \nonumber
\\ && \label{34}
\end{eqnarray}
The completeness relation may be written in a more compact form in terms of the matrices
themselves rather than the column vectors, i.e.,
\begin{eqnarray}
\sum_{j=1}^{n_b} \psi_{E_j}(x) \psi_{E_j}^\dagger(x') & + & 
\frac{1}{2\pi} \int_0^\infty dk  \; [ \psi(k,x)
\psi^\dagger(k,x') \nonumber \\ 
& + & \widetilde{\psi}(k,x)
\widetilde{\psi}^\dagger(k,x')] = {\bf 1} \delta(x-x'). \label{35}
\end{eqnarray}
For the free-particle case when $V(x)={\bf 0}$, there are no bound
states and the completeness relation is
\begin{eqnarray}
&\displaystyle{\frac{1}{2\pi} \int_0^\infty \; dk [ \psi^0(k,x)
\psi^{0\dagger}(k,x') + \widetilde{\psi}^0(k,x)
\widetilde{\psi}^{0\dagger}(k,x')]} & \nonumber \\
&  = {\bf 1} \delta(x-x'). & \label{36}
\end{eqnarray}
We now subtract Eq.~(\ref{36}) from Eq.~(\ref{35}), set $x'=x$, and
integrate over $x$ from $-{\cal R}$ to ${\cal R}$.  The resulting equation may be
written as
\begin{eqnarray}
 & {\displaystyle \int_0^\infty dk \int_{-{\cal R}}^{\cal R} dx \; [ \psi(k,x)
\psi^\dagger(k,x) + \widetilde{\psi}(k,x)
\widetilde{\psi}^\dagger(k,x)} & \nonumber \\
& {\displaystyle {- \psi^0(k,x)
\psi^{0\dagger}(k,x) - \widetilde{\psi}^0(k,x)
\widetilde{\psi}^{0\dagger}(k,x)]}} & \nonumber \\
&{\displaystyle  = - 2\pi \int_{-{\cal R}}^{\cal R} dx 
\sum_{j=1}^{n_b} \psi_{E_j}(x) \psi_{E_j}^\dagger(x).} &  \label{37}
\end{eqnarray}
The trace of Eq.~(\ref{37}) in the limit as ${\cal R}$ approaches infinity
gives 
\begin{eqnarray}
&&  \displaystyle{\lim_{{\cal R}\rightarrow \infty} 
\int_0^\infty dk \int_{-{\cal R}}^{\cal R} dx \; \mbox{Tr}[ \psi(k,x)
\psi^\dagger(k,x) +  \widetilde{\psi}(k,x)
\widetilde{\psi}^\dagger(k,x)}  \nonumber \\
&& \displaystyle{ - \psi^0(k,x)
\psi^{0\dagger}(k,x) - \widetilde{\psi}^0(k,x)
\widetilde{\psi}^{0\dagger}(k,x)] = -2\pi n_b}  \label{38}
\end{eqnarray}
To perform the integration over $x$ in the right side of
Eq.~(\ref{38}), we use the identity
\begin{equation}
\mbox{Tr}[\psi\psi^\dagger] = \frac{1}{2k} \frac{\partial~}{\partial x} 
\left\{ \mbox{Tr}\left[ \frac{\partial \psi}{\partial k} 
\frac{\partial\psi^\dagger}{\partial x} - \frac{\partial^2 \psi}
{\partial x \partial k} \psi^\dagger \right] \right\},  \label{39}
\end{equation}
which may be obtained by taking the derivative with respect to $k$ of
the Schr\"{o}dinger equation (\ref{1}).  Since in the limit ${\cal R}$
exceeds the range of the potential, we can insert the asymptotic forms
of the wave functions in Eq.~(\ref{38}) to obtain
\begin{eqnarray}
& & \displaystyle{\lim_{{\cal R}\rightarrow\infty} \int_0^\infty 
\frac{dk}{2k} \; \mbox{Tr}\left[-2ik\left(\frac{\partial\tau}{\partial k}
\tau^\dagger + \frac{\partial\rho}{\partial k} \rho^\dagger + 
\frac{\partial\widetilde{\tau}}{\partial k}
\widetilde{\tau}^\dagger + \frac{\partial\widetilde{\rho}}{\partial k} 
\widetilde{\rho}^\dagger \right) \right.}  \nonumber \\
&&  \displaystyle{\left. - i (\rho + \widetilde{\rho})e^{2ik{\cal R}} +
i(\rho^\dagger + \widetilde{\rho}^\dagger)e^{-2ik{\cal
R}}\frac{}{}\right] = -2\pi n_b.}  \label{40}
\end{eqnarray}
Following Newton and Jost\cite{Newton55} we define the phase as
\begin{equation}
\eta(k) = \frac{1}{2i} \ln \det S(k),   \label{41}
\end{equation}
where we require $\eta(k)$ to be continuous  for $k \in (0,\infty)$.
Since the $S$ matrix is unitary, we may write $S=U^\dagger S_DU$ where
$U$ is a real orthogonal matrix and $S_D$ is the
diagonal matrix $S_D =
\mbox{diag}(e^{i\delta_1},\dots,e^{i\delta_{2N}})$ where the $\delta_j$'s
are real phases~\cite{Mott65}.  Let us write therefore $S=e^{i\Delta}$ where $\Delta
= U^\dagger\Delta_DU$ for $\Delta_D =
\mbox{diag}(\delta_1,\dots,\delta_{2N})$.  Then
\begin{eqnarray}
2i\eta(k) & = & \ln\det S(k) = \ln\det S_D(k) \nonumber \\
& = &\displaystyle{\sum_{j=1}^{2N} i\delta_j(k)} =
\mbox{Tr}[i\Delta_D(k)]
\end{eqnarray}
and
\begin{eqnarray}
2i\frac{\partial \eta}{\partial k} & = & \mbox{Tr}\left[\frac{\partial
i\Delta_D}{\partial k}\right]=\mbox{Tr}\left[S^\dagger_D\frac{\partial
S_D}{\partial k}\right] = \mbox{Tr}\left[S^\dagger\frac{\partial
S}{\partial k}\right] \nonumber \\
 & = & \mbox{Tr} \left[ \frac{\partial \tau}{\partial k} \tau^\dagger
+\frac{\partial\rho}{\partial k} \rho^\dagger + \frac{\partial 
\widetilde{\tau}}{\partial k} \widetilde{\tau}^\dagger
+\frac{\partial\widetilde{\rho}}{\partial k} \widetilde{\rho}^\dagger 
\right].  \label{43}
\end{eqnarray} 
Thus Eq.~(\ref{40}) may be written as
\begin{equation}
\eta(0)-\eta(\infty) = \pi n_b - \lim_{{\cal R}\rightarrow\infty} X({\cal R}),
\label{44}
\end{equation}
where 
\begin{eqnarray}
X({\cal R}) & = & \frac{1}{2} \int^\infty_0 \frac{dk}{2k} \; \mbox{Tr}\left[ 
i(\rho(k)\right.  + \widetilde{\rho}(k)) e^{2ik{\cal R}} \nonumber \\
& & \left. - i(\rho^\dagger(k) 
+\widetilde{\rho}^\dagger(k)) e^{-2ik{\cal R}} \right].  \label{45}  \\
 & = & \frac{i}{4} \int^\infty_{-\infty} \frac{dk}{k} \;
\mbox{Tr}[\rho(k) + \widetilde{\rho}(k)]e^{2ik{\cal R}}.  \label{45a}
\end{eqnarray}
In the next section we show that $\rho(k) \sim O(1/k)$ for large
$|k|$, so that the integration in
Eq.~(\ref{45a}) converges for large $|k|$.  We now take the limit as
${\cal R}$
approaches $\infty$, using the relation 
\begin{equation} 
\lim_{{\cal R}\rightarrow\infty} \frac{e^{2ik{\cal R}}}{k} = i\pi\delta(k),  
\end{equation}
where $\delta(k)$ is the Dirac delta function.
In that limit $X({\cal R})$ goes to 
 $-(\pi/4)\mbox{Tr} [\rho(0) + \widetilde{\rho}(0)]$, so
that the statement of Levinson's theorem now is
\begin{equation}
\eta(0) = \pi n_b + \frac{\pi}{4}
\mbox{Tr}[\rho(0)+\widetilde{\rho}(0)], \label{48}
\end{equation}
where we have set $\eta(\infty)$ equal to zero.  In the next
section we also show that
\begin{equation}
\mbox{Tr}[\rho(0) +\widetilde{\rho}(0)] = -2(N-n),  \label{49}
\end{equation}
where $n$ is the number of 
``half-bound states''~\cite{Newton80,Newton83,Aktosun85}.
Thus in its final form Levinson's theorem states,
\begin{equation}
\eta(0) = \pi(n_b + \frac{1}{2} n - \frac{1}{2} N). \label{50}
\end{equation}
This expression of the theorem is consistent with that for the
uncoupled case given in Ref.~\cite{Newton83}.

\subsection{Threshold behaviour of $\rho$ and $\tau$}
\label{threshold}

The threshold behaviour of reflection and transmission amplitudes and
coefficients has been discussed recently in
several articles~\cite{DeBianchi94,Senn88,VanDijk92,Nogami95}.  In
order to study the behaviour of the $\rho$ and $\tau$ matrices
at $k=0$, we introduce a different set of solutions of the
Schr\"{o}dinger equation, since according to Eq.~(\ref{3b}) the
columns of $\psi$ and
$\widetilde{\psi}$ fail to be linearly independent at $k=0$.
Let $\phi(k,x)$ and $\chi(k,x)$ be solution
matrices of Eq.~(\ref{1}) which
satisfy the boundary conditions,
\begin{equation}
\begin{array}{lll}
\phi(k,-R) & = \chi'(k,-R) & = \bf{1} \\
\phi'(k,-R) & = \chi(k,-R) & = \bf{0},
\end{array} \label{51}
\end{equation}
where $R$ is the range of the potential as defined at the beginning of
Sec. II.  By evaluating $\det{W(\phi,\chi)}$ at $x=-R$ and using the 
results of Appendix A, we readily show that
for all $x$ and $k$ the $\det{W(\phi,\chi)}=1$, where the matrix $W$ is
defined as in Eq.~(\ref{3a}).  Thus unlike the column vectors of
$\psi$ and $\widetilde{\psi}$, the column vectors of $\phi$ and $\chi$
are linearly independent at zero energy.

In order to obtain the scattering amplitudes,  we
expand $\psi$ in terms of $\phi$ and $\chi$ so that
\begin{equation}
\psi(k,x) = \phi(k,x) B(k) + \chi(k,x) C(k), \label{52}
\end{equation}
where $B(k)$ and $C(k)$ are matrices of expansion coefficients.
Evaluating $\psi$ and $\psi'$ at $\pm R$ using Eqs.~(\ref{2}) and 
(\ref{52}), we obtain four equations involving $B(k), C(k),
\rho(k)$ and $\tau(k)$.  By eliminating three of these we find that
$\rho$ can expressed in terms of $\phi$ and $\chi$ and their
derivatives evaluated at $R$.  Thus
\begin{eqnarray}
&&\rho(k)  =  \{k^2 \chi(k,R) +ik[\chi'(k,R) + \phi(k,R)]
- \phi'(k,R)\}^{-1} \nonumber \\
 & &\{k^2 \chi(k,R) +ik[\chi'(k,R) - \phi(k,R)] +
\phi'(k,R)\} e^{-2ikR}. \nonumber \\  \label{53}
\end{eqnarray}
Similarly by expanding $\widetilde{\psi}$ in terms of $\phi$ and $\chi$ we
obtain 
\begin{eqnarray}
&&\widetilde{\rho}(k) = \{k^2 \chi(k,R) - ik[\chi'(k,R) - \phi(k,R)] +
\phi'(k,R)\} \nonumber \\
 & &\{k^2 \chi(k,R) +ik[\chi'(k,R) + \phi(k,R)] -
\phi'(k,R)\}^{-1} e^{-2ikR},  \nonumber \\ \label{54}
\end{eqnarray}
\begin{eqnarray}
\widetilde{\tau}(k) & = & 2ik \{k^2\chi(k,R) + ik[\chi'(k,R) +
\phi(k,R)] \nonumber \\
&& - \phi'(k,R)\}^{-1} e^{-2ikR},  \label{65}
\end{eqnarray}
and, since $\tau(k)=\widetilde{\tau}^T(k)$,
\begin{eqnarray}
 \tau(k) & = & 2ik\{k^2\chi^T(k,R) + ik[\chi'^T(k,R) + \phi^T(k,R)]
\nonumber \\
&& - \phi'^T(k,R)\}^{-1} e^{-2ikR}.  \label{54a}
\end{eqnarray} 
Thus all four scattering amplitudes, and consequently the $S$ matrix,
are determined by $\phi$ and $\chi$ and their derivatives evaluated at
$R$.  In Appendix B it is shown that $\phi(k,x)$ and $\chi(k,x)$ are
entire functions of complex $k$ for all $x\in [-R,R]$ so that the analytic
properties of the scattering amplitudes can be determined using
these wave functions.  By inserting the expressions for the large
real $k$ behaviour of the wave functions, Eqs.~(\ref{A210}) and
(\ref{A211}), into the expressions for the scattering amplitudes, we
obtain
\begin{equation}
\tau(k) \sim {\bf 1} + O(1/k)\mbox{ and } 
\rho(k) \sim {\bf 0} + O(1/k)  \mbox{ for } k
\rightarrow \infty,
\end{equation}
and similar expressions for $\widetilde{\tau}$ and $\widetilde{\rho}$.

If $\det[\phi'(0,R)] \neq 0$, the reflection and
transmission amplitudes at zero energy are
\begin{equation}
\rho(0)  = \widetilde{\rho}(0) = -{\bf 1} \hspace{.2in} \mbox{and}
\hspace{.2in} \tau(0)  = \widetilde{\tau}(0)  = {\bf 0}.
 \label{55}
\end{equation}
The case for which $\det[\phi'(0,R)] = 0$ needs special
attention.  In order to understand the significance of this condition,
we look at the bound states of the system.  In the Schr\"{o}dinger
equation~(\ref{1}) for bound states, we denote the bound-state
energy as $\alpha^2=-k^2$ with $\alpha >0$.  The bound-state wave
functions can be expressed as column vectors of a
matrix $\psi_b(\alpha,x)$ with the asymptotic boundary conditions,
\begin{equation}
\psi_b(\alpha,x) \sim \cases{ e^{\alpha x}Q, & $x \leq -R$ \cr
e^{-\alpha x} T,  & $x \geq R$\cr} \label{56}
\end{equation}
where Q and T are matrices of constants.  The number of independent
bound states at a given energy will depend on the rank of the matrix
$\psi_b$, and consequently cannot exceed $N$. Proceeding as we did for the
scattering states, we expand the bound-state wave functions in terms
of the functions $\phi_b(\alpha,x)$ and $\chi_b(\alpha,x)$ which are
solution matrices of the Schr\"{o}dinger equation with energy $-\alpha^2$ and
satisfy the boundary conditions
\begin{equation}
\begin{array}{lll}
\phi_b(\alpha,-R) & = \chi'_b(\alpha,-R) & = \bf{1}  \\
\chi_b(\alpha,-R) & = \phi'_b(\alpha,-R) & = \bf{0}.
 \end{array}  \label{57}
\end{equation}
Thus
\begin{equation}
\psi_b(\alpha,x) = \phi_b(\alpha,x) \beta(\alpha) + \chi_b(\alpha,x)
\gamma(\alpha), \label{58}
\end{equation}
where $\beta$ and $\gamma$ are matrices of expansion coefficients.  At
$R$ and $-R$ we
match the asymptotic form of the wave function, Eq.~(\ref{56}), and
its derivative to the expanded form, Eq.~(\ref{58}), and its
derivative.  Eliminating $T, Q$ and $\gamma$ from the
four equations so obtained, we are left with the equation
\begin{eqnarray}
& \{\alpha^2 \chi_b(\alpha,R) + \alpha[\chi'_b(\alpha,R) +
\phi_b(\alpha,R)] & \nonumber \\
& + \phi'_b(\alpha,R)\} \beta(\alpha) = 0 & \label{59} 
\end{eqnarray}
Since one of the four matching equations is $Q=\beta(\alpha)e^{\alpha R}$,
there will be bound states only if the matrix $\beta(\alpha)$ contains
nonzero entries.  Such a nontrivial matrix exists only when
\begin{eqnarray}
& \det\{\alpha^2 \chi_b(\alpha,R) + \alpha[\chi'_b(\alpha,R) +
\phi_b(\alpha,R)] & \nonumber \\ 
& + \phi'_b(\alpha,R)\} = 0. & \label{60}
\end{eqnarray}
This is the bound-state eigenvalue equation for energy $-\alpha^2$.
In contrast to the case with no coupling, the bound-state eigenenergies
can be degenerate.

Let us consider the eigenstates when $\alpha = 0$.  These will occur
 only if $\det[\phi_b'(0,R)]=0$.  In general the solutions represented
 by the columns vectors of $\psi_b(0,x)$ are bounded but not square
 integrable; hence they are referred to as half-bound
 states~\cite{Newton77}.  The restriction on the potential function
 that it vanishes for $|x|\geq R$, precludes the possibility of having
 normalizable state functions at zero energy.  For this to be the case
 a linear combination of the columns of $\psi_b$ would yield
 $\Psi(0,x)=0$ for $|x|\geq R$.  Such a boundary condition would lead
 to the trivial solution of Eq.~(\ref{1}).  Normalizable zero-energy bound
 states can exist for potentials which are less restrictive than those
of this paper~\cite{Aktosun85}.

Since $\phi(0,x)$ and $\phi_b(0,x)$ are solutions of the same system
of differential equations and both have the same boundary conditions,
$\phi_b(0,x) = \phi(0,x)$.
Thus the condition that
\begin{equation}
\det[\phi'(0,R)]=0  \label{62}
\end{equation}
is equivalent to the condition for the existence of half-bound states.

Consider the matrix eigenvalue equation,  
\begin{equation} 
\phi'(0,R) \widetilde{\beta} = \lambda \widetilde{\beta},   \label{63}
\end{equation} 
where $\widetilde{\beta}$ is a column vector
and $\lambda$ is its eigenvalue.  There will be a nontrivial solution only if
\begin{equation}
\det[\phi'(0,R)-\lambda {\bf 1}] = 0.  \label{64}
\end{equation}
Suppose the eigenvalues obtained are $\lambda_1,\cdots,\lambda_N$.  At
least one of these must be zero if there is a half-bound
state.  Actually there may be $n(\leq N)$ zero eigenvalues.  We can
order these in the following way:
$0,\dots,0,\lambda_{n+1},\dots,\lambda_N$.  These $n$ zero eigenvalues
will have $n$ linearly independent eigenvectors associated with them,
which represent $n$ distinct half-bound states.  

We now return to the discussion of reflection and transmission amplitudes.
The inverse of $\widetilde{\tau}$ of Eq.~(\ref{65}) can be written as
\begin{eqnarray}
2ik\widetilde{\tau}^{-1}(k) & = & \{k^2\chi(k,R) + ik[\chi'(k,R) +
\phi(k,R)] \nonumber \\
&& - \phi'(k,R)\}e^{2ikR}.  \label{66}
\end{eqnarray}
If this equation is combined with Eq.~(\ref{54}), we obtain
\begin{eqnarray}
2ik\widetilde{\rho}(k)\widetilde{\tau}^{-1}(k) & = & \{k^2\chi(k,R)
-ik[\chi'(k,R) - \phi(k,R)] \nonumber \\ 
&& + \phi'(k,R)\}.  \label{67} 
\end{eqnarray}
The factor in the curly brackets of Eq.~(\ref{66}) is precisely that
in the determinant of Eq.~(\ref{60}) (with $k$$=$$i\alpha$), i.e., it
is the factor which determines the bound states of the system.  
In Appendix B we show that matrices $\phi$ and $\chi$ are elementwise
entire functions of $k$.  From Eq.~(\ref{66}) and the fact that
$\widetilde{\tau}^T = \tau$ we see that $2ik\tau^{-1}(k)$ and $(2ik)^N
\det{\tau^{-1}(k)}$ are also entire functions of $k$.  According to
Eqs.~(\ref{60}) and (\ref{66}) $\det{\tau^{-1}(k)}$ has a zero at
$k=i\alpha$ when $-\alpha^2$ is the energy of a bound state.  Thus
$\det{\tau(k)}$ has poles (possibly multiple) on the positive
imaginary axis of the complex $k$ plane.  In the absence of a
half-bound state, $\det{\tau^{-1}(k)}$ has an $N^{\rm th}$ order pole
at $k=0$ and $\det{\tau}$ has an $N^{\rm th}$ order zero at $k=0$.
Since there can be no more than $N$ half-bound states $\det{\tau(k)}$ and
$\tau(k)$ are analytic in a neighborhood of $k=0$.

In the following we consider real $k$. Taking the limits as $k$ goes
to zero of Eqs.~(\ref{66}) and (\ref{67}) we obtain
\begin{eqnarray}
{\displaystyle\lim_{k\rightarrow 0}}\;\; 2ik \widetilde{\tau}^{-1}(k)  
& = & -\phi'(0,R)   \mbox{ and } \nonumber \\
{\displaystyle \lim_{k\rightarrow 0}} \;\; 2ik \widetilde{\rho}(k)
\widetilde{\tau}^{-1}(k) & = & \phi'(0,R).  \label{68}
\end{eqnarray}
Similarly using Eq.~(\ref{54a}) and the transpose of Eq.~(\ref{53}), we
get
\begin{eqnarray}
{\displaystyle\lim_{k\rightarrow 0}}\;\; 2ik \tau^{-1}(k)  
& = & -\phi'^T(0,R) \mbox{ and } \nonumber \\ 
{\displaystyle \lim_{k\rightarrow 0}} \;\; 2ik \rho(k)
\tau^{-1}(k) & = & \phi'^T(0,R).  \label{68a}
\end{eqnarray}
We introduce unitary matrices $U(k)$ and $V(k)$ which diagonalize
$\tau$~\cite[page 192]{Lancaster85}, so that
\begin{eqnarray}
\tau_D(k) & = & U^\dagger(k)\tau(k)V(k) \mbox{ and} \nonumber \\
 2ik\tau_D^{-1} & = & V^\dagger(k) 2ik\tau^{-1}(k) U(k).  \label{69}
\end{eqnarray}
In the limit as $k$ approaches zero $U(0)$ and $V(0)$ also diagonalize
 $\phi'^T(0,R)$, i.e., $\phi'^T_D(0,R) = V^\dagger(0)
\phi'^T(0,R) U(0)$, so that
\begin{equation}
\lim_{k\rightarrow 0} \; 2ik\tau_D^{-1}(k) = -{\phi'_D}^T(0,R).
\label{70}
\end{equation}
We define the matrix
\begin{equation}
r(k) \equiv V^\dagger(k)\rho(k)V(k), \label{71}
\end{equation}
so that
\begin{equation}
\lim_{k\rightarrow 0} \; 2ik r(k) {\tau_D}^{-1}(k) 
= {\phi'_D}^T(0,R).   \label{72}
\end{equation}
Combining Eqs.~(\ref{70}) and (\ref{72}) gives
\begin{equation}
r(0) {\phi'_D}^T(0,R) = -{\phi'_D}^T(0,R).  \label{73}
\end{equation}
The matrices $\phi'^T(0,R)$ and ${\phi'_D}^T(0,R)$ have the same
rank~\cite[page 55]{Lancaster85}.  Thus ${\phi'_D}^T(0,R)$ will have the
same number of nonzero diagonal elements as there are nonzero
eigenvalues of $\phi'^T(0,R)$ or $\phi'(0,R)$.  
Writing ${\phi'_D}^T(0,R) =
\mbox{diag}(0,\cdots,0,s_{n+1},\cdots,s_N)$ and using
Eq.~(\ref{73}) we see that the matrix $r(0)$ must have the
form
\begin{equation}
r(0) = \left( \begin{array}{cc} R_{11} & {\bf 0}  \\ R_{21} &
-{\bf 1} \end{array}
\right),  \label{74}
\end{equation}
where the matrices $R_{11}$, $R_{21}$, ${\bf 0}$ and ${\bf 1}$ have
dimensions $n\times n$, $(N-n)\times n$, $n\times(N-n)$ and
$(N-n)\times (N-n)$ respectively. 

To study the behaviour of $\tau_D(k)$ near $k=0$, we consider the
Hermitian positive semi-definite matrix $T(k)=\tau^\dagger(k)\tau(k)$,
whose real nonnegative eigenvalues we denote by
$t_1^2(k),\dots,t_N^2(k)$.  The singular values of $\tau(k)$
are defined as the nonnegative square root of these, i.e.,
$t_1(k),\dots,t_N(k)$, and they form the diagonal elements of
$\tau_D(k)$ in Eq.~(\ref{69}), so that $\tau_D(k) =
\mbox{diag}(t_1(k),\dots,t_N(k))$~\cite[page 192]{Lancaster85}.  Since
$\tau(k)$ is analytic in the neighborhood of $k=0$, so is $T(k)$.  For
small real $k>0$, we invoke a theorem of Rellich~\cite[page 31]{Rellich69}
which states that the eigenvalues of $T(k)$ are convergent power
series of $k$.  Furthermore, using Eq.~(\ref{27a}), we note that $T(k)
= T^T(-k)$ and the eigenvalues of $T(k)$ and $T(-k)$ are the same.
Thus taking the eigenvalues in the same order, we find that
$t_i^2(-k) = t_i^2(k)$.  Hence the $t_i^2(k)$'s are power series of
$k^2$ and the $t_i(k)$'s are power series of $k$.  Thus we may write
$\tau_D(k)= \mbox{diag}(t_1+t_{11}k+\cdots,\dots, t_N+t_{N1}k+\cdots)$
and $\tau_D^{-1}(k)= \mbox{diag} ((t_1+t_{11}k+\cdots)^{-1} , \dots
,(t_N+t_{N1}k+\cdots)^{-1})$, where the $t_i$'s and $t_{ij}$'s are
constants.  In order that
\begin{eqnarray}
\lim_{k\rightarrow 0} 2ik\tau_D^{-1}(k) & = & \lim_{k\rightarrow 0}
\mbox{diag} (2ik(t_1+t_{11}k+\cdots)^{-1}, \dots, \nonumber \\
&& \hspace{.2in} 2ik(t_N + t_{N1} k + \cdots)^{-1}) \\
& = & \mbox{diag}(0,\dots,0,-s_{n+1},\dots,-s_N),
\end{eqnarray}
the quantities $t_1,\dots,t_n \neq 0$ and $t_{n+1}=\cdots
=t_N = 0$.  The matrix $\tau_D(0)$ therefore has the form
$\tau_D(0)=\mbox{diag}(t_1,\dots ,t_n,0,\dots,0)$ where the
first $n$ diagonal elements are nonzero.

Recall that $\tau^T=\widetilde{\tau}$ and therefore $(\tau^T)^{-1} =
\widetilde{\tau}^{-1}$, and that
$\widetilde{\rho}=\widetilde{\rho}^T$.  The second part of
Eq.~(\ref{68}) becomes
\begin{equation}
\lim_{k\rightarrow 0} \; \; 2ik \tau^{-1}(k) \widetilde{\rho}(k) =
\phi'^T(0,R). \label{75a}
\end{equation}
The same $U$ and $V$ can be used to obtain
\begin{equation}
\lim_{k\rightarrow 0} \; \; 2ik {\tau_D}^{-1}(k) \widetilde{r}(k) =
{\phi'_D}^T(0,R), \label{75b}
\end{equation}
where
\begin{equation}
\widetilde{r}(k) = U^\dagger(k)\widetilde{\rho}(k)U(k). \label{75c}
\end{equation}
As before the structure of $\widetilde{r}(0)$ may be determined, and
it is
\begin{equation}
\widetilde{r}(0) = \left( \begin{array}{cc} \widetilde{R}_{11} &
\widetilde{R}_{12}  \\  {\bf 0} & -{\bf 1} \end{array}
\right).  \label{76}
\end{equation}
In order to simplify Eq.~(\ref{48}) we write
\begin{eqnarray}
\mbox{Tr}[\rho(0)+\widetilde{\rho}(0)] & = &
\mbox{Tr}[r(0)+\widetilde{r}(0)] \nonumber  \\
& = & -2(N-n) + \mbox{Tr}[R_{11}^{n\times n} +
\widetilde{R}_{11}^{n\times n}],  \label{78}
\end{eqnarray}
where the superscripts refer to the dimensions of the matrices.
All that remains is to evaluate the last term of the right side of this
equation.  At zero energy the scattering amplitudes are real.  Using
this fact along with the relations that $\widetilde{\tau}=\tau^T$ and
$\rho=\rho^T$ in Eq.~(\ref{15}), we obtain
\begin{equation}
\tau(0)\rho(0)+\widetilde{\rho}(0)\tau(0)={\bf 0}  \label{79}
\end{equation}
By applying the diagonalization transformation of $\tau$, we
find that
\begin{equation}
\tau_D(0)r(0) + \widetilde{r}(0)\tau_D(0) = {\bf 0}.  \label{80}
\end{equation}
In matrix form this equation may be written
\begin{equation}
\left(\begin{array}{cc} A_1 & {\bf 0} \\ {\bf 0} & {\bf 0} \end{array}
\right) \left( \begin{array}{cc} R_{11} & {\bf 0} \\ R_{21}  & -{\bf 1}
\end{array} \right) + \left( \begin{array}{cc} \widetilde{R}_{11} &
\widetilde{R}_{12} \\ {\bf 0} & -{\bf 1} \end{array} \right) \left(
\begin{array}{cc} A_1 & {\bf 0} \\ {\bf 0} & {\bf 0} \end{array}
\right) = {\bf 0}, \label{81}
\end{equation}
where $A_1 = \mbox{diag}(t_1,\cdots,t_n)$ and the block matrices
have the appropriate dimensions.  From Eq.~(\ref{81}) it follows that
\begin{equation}
A_1 R_{11} + \widetilde{R}_{11} A_1 = {\bf 0}^{n\times n}.  \label{82}
\end{equation}
Consequently, the second term in the right side of Eq.~(\ref{78}) is
zero, and we have the simple relation
\begin{equation} 
\mbox{Tr}[\rho(0)+\widetilde{\rho}(0)] = -2(N-n) \label{83}
\end{equation}

In light of the discussion of the $N=1$ normal and anomalous threshold
behaviour~\cite{Senn88,Nogami95}, our result seems surprising since
the right side of Eq.~(\ref{83}) is an integer, whereas the anomalous
threshold effect for parity-noninvariant potentials gives values for
$\rho(0)$ which lie between 0 and 1.  However, if one considers the
sum of $\rho(0)$ and $\widetilde{\rho}(0)$ in the uncoupled case,
one obtains an even integer.  For a parity-invariant potential, i.e.,
$V(-x)= V(x)$, $\rho$ and $\widetilde{\rho}$ are equal and the
zero-energy value of $\rho$ will always be an integer.


\section{Segmented potentials and factorization of the $S$ matrix}
\label{segment}

For the Schr\"{o}dinger equation without coupling it is well known
that the reflection and transmission amplitudes satisfy a
factorization formula.  That is, if the potential is subdivided into a
number of sections, then the total transmission and reflection
amplitudes for the system can be expressed in terms of the amplitudes
for each of the truncated pieces of the potential.  The recent proof
of Aktosun~\cite{Aktosun92} may be generalized immediately to the
case of $N$ coupled equations.

Following Aktosun, then, we subdivide the real line into $J$ pieces.
The boundaries of the segments are denoted by $x_i$, $i$$=$$0,\ldots,
J$, with $-R=x_0$$<$$x_1$$<\ldots<$$x_{J-1}$$<$$x_J=R$.  The potential
may then be written as a sum of truncated potentials as follows
\begin{equation}
 V(x)=\sum_{j=0}^{J-1}V_{j}(x), 
\end{equation}
 where 
\begin{equation}
V_{j}(x)=\cases{V(x), & $x_j<x<x_{j+1}$\cr
 0, &  otherwise.\cr}
\end{equation}
 The single indices on the truncated potentials should not
be confused with the implicit double indices which label the various
elements of the potential matrix.

For a given $j$, then, let us define the matrix 
\begin{equation}
	\Lambda_j(k)=\left(\begin{array}{cc}
		  \tau_j^{-1}(k) & -\tau_{j}^{-1}(k){\widetilde \rho}_j(k)\\
		  \rho_j(k)\tau_j^{-1}(k) & \left({\widetilde \tau}_j^{\dagger}
		      (k)\right)^{-1}
		\end{array}\right),
\end{equation}
where $\tau_j(k),\widetilde{\tau}_j(k),\rho_j(k),{\widetilde
\rho}_j(k)$ 
represent the various amplitude
matrices for the truncated potentials with the usual boundary conditions.
The amplitudes for the original potential are similarly arranged into
a matrix,
\equation 
        \Lambda(k)=\left(\begin{array}{cc} 
                  \tau^{-1}(k) & -\tau^{-1}(k){\widetilde \rho}(k)\\
                  \rho(k)\tau^{-1}(k) & \left({\widetilde \tau}^{\dagger}
                      (k)\right)^{-1}  
		\end{array}\right) ,
\endequation
and the factorization formula is then simply given by
\equation
	\Lambda(k)=\prod_{j=0}^{J-1}\Lambda_j(k),
	\label{facteq}
\endequation
where the factors on the right side of the equation are ordered so
that factors with lower subscripts occur to the left of the ones with
higher subscripts.
The proof of Eq.~(\ref{facteq}) is completely analogous to that
advanced by Aktosun in the $N=1$ case and so we shall not review
it here.  The only added complication is the fact that
the various amplitude matrices do not generally commute, but this has been 
properly accounted for in the definitions of the $\Lambda$ matrices.
The utility of Eq.~(\ref{facteq}) will become apparent below when we use
it to derive the amplitudes for scattering from two different 
delta-function matrix potentials in terms of the amplitudes for scattering
from each of them separately.  

This approach effectively factorizes the $S$ matrix in the sense that
if the $S$ matrix of the $j^{\rm th}$ potential segment, i.e., 
\begin{equation}
S_j(k) = \left( \begin{array}{cc} 
 		~\tau_j(k)~ & ~\widetilde{\rho}_j(k) \\
		\rho_j(k)   & ~\widetilde{\tau}_j(k)
		\end{array}
    	\right),
\end{equation}
is known, then the scattering amplitudes of the potential segment are
determined, and from them $\Lambda_j(k)$.  Using Eq.~(\ref{facteq}) we can
obtain $\Lambda(k)$ for the whole potential, and this allows us to solve
for the scattering amplitudes and the $S$ matrix of the whole potential.  

Another generalization of the uncoupled to the coupled
problem involves the finitely periodic potentials, recently discussed
by a number of
authors~\cite{DeBianchi95,Rozman94a,Rozman94b,Sprung93}. In the
derivation of the factorization formula, Eq.~(\ref{facteq}), we have
to be careful in the ordering of the products such as $\rho
\tau^{-1}$.  
This non-commutativity of the amplitude matrices would typically
prevent us from generalizing the closed-form solutions of the
finitely periodic potentials.  There are however classes of potentials
for which the various amplitude matrices do commute with each other
and for which the results for no coupling {\em can} be generalized.  

An example of such a class of potentials consists of those potentials
which can be expressed as 
\begin{equation} 
V(x) = U \mbox{diag}[v_1(x),v_2(x),\dots,v_N(x)] U^T,
 \label{potsc}
\end{equation} 
where $U$ is a constant (real) orthogonal matrix.
Note that the potentials of Eq.~(\ref{potsc}) include as a subclass
those of the form $V(x)=v(x) M$, where $v(x)$ is
a real-valued function of $x$ and $M$ is a constant symmetric matrix.
Such potentials have been used previously in various applications
(see, for example, Ref.~\cite{Doyle75}).  It is easy to prove
that when the potential is diagonalizable by a constant orthogonal
matrix, then all of the amplitude matrices (as well as their inverses
and hermitian conjugates) commute with each other.

If we use potentials of this type to construct finitely periodic
potentials with nonoverlapping subpotentials, the analysis of Rozman
{\em et al.}\cite{Rozman94a,Rozman94b} follows in the same way for the
matrix potential problem and the expressions for the amplitude matrices are
straight forward generalizations of their results.

\section{Potential Models}
\label{model}
Below we consider some potential models which lend themselves to
solutions in closed form.  These models help to elucidate 
some of the results obtained in the previous sections.

\subsection{Constant potential matrix}

An example of a potential for which solutions can be
obtained in closed form is the 
square-well or square-barrier potential matrix for which
\begin{equation}
V(x)= \cases{V_0  &  for $a\leq x\leq b$ where $a\geq -R$ and $b\leq R$\cr
 0 & otherwise,\cr}  \label{100}
\end{equation}
where $V_0$ is a real symmetric $N\times N$ matrix.
The Schr\"{o}dinger equation~(\ref{1}) is equivalent to a
first-order differential equation of the matrix function
$W(x)=W(\phi,\chi)$ of Eq.~(\ref{A2}), i.e., 
\begin{equation}
W'(x)= F(x) W(x),   \label{51b}
\end{equation}
where
\begin{equation}
F(x) = \left( \begin{array}{cc} {\bf 0} & {\bf ~~1~~} \\
                                  V(x)-k^2 & {\bf 0}
              \end{array} \right), \label{100a}
\end{equation}
with the boundary condition $W(-R)=I$.  As we saw earlier the function
$W(x)$ provides the advantage of giving solutions that are
linearly independent at $k=0$.  In addition, Eq.~(\ref{51b}) 
gives us an initial
value problem, rather than the two-point boundary condition problem of
the original Schr\"{o}dinger equation.  Solving for the scattering
amplitudes numerically is simpler for the initial value
problem.  An alternative to this approach is the variable amplitude
formulation which also casts the problem into a system of first order
differential equations with an initial value
condition~\cite{VanDijk79,VanDijk79a}.  In principle, Eq.~(\ref{51b})
can be used to solve the Schr\"{o}dinger equation for any arbitrary
potential matrix.

For the constant potential matrix the differential equation Eq.~(\ref{51b}) can
be integrated starting at $x=-R$ over the three regions $(-R,a),
(a,b)$ and $(b,R)$ in turn\cite{Henrici77}.  The result is
\begin{eqnarray}
W(R) & = & \left( \begin{array}{cc} \cos k(R-b) & k^{-1}\sin k(R-b) \\
                                -k \sin k(R-b)  & \cos k(R-b) 
              \end{array}  \right) \nonumber \\
&&\hspace{.1in} \left( \begin{array}{cc} \cosh K(b-a) & K^{-1}\sinh K(b-a) \\
                                K \sinh K(b-a)  & \cosh K(b-a) 
              \end{array}  \right) \nonumber \\ 
& &\hspace{.2in} \left( \begin{array}{cc} \cos k(a+R) & k^{-1}\sin k(a+R) \\
                                -k \sin k(a+R)  & \cos k(a+R) 
              \end{array}  \right),  \label{101}
\end{eqnarray}
where $K^2 = V_0 - k^2$.  To simplify matters, but still to allow us
to study a model with a potential function without definite parity, we set $a =
-R$, so that we obtain explicit forms for the wave functions at $x=R$.
\begin{equation}
\begin{array}{lll}
\phi(k,R) & = & \cos k(R-b)\cosh K(R+b) \\ 
&&\hspace{.2in} + k^{-1} \sin k(R-b) K \sinh K(R+b) \\
\phi'(k,R) & = & -k \sin k(R-b) \cosh K(R+b) \\
&&\hspace{.2in} + \cos k(R-b) K \sinh K(R+b) \\
\chi(k,R) & = & \cos k(R-b) K^{-1} \sinh K(R+b) \\
&&\hspace{.2in} + k^{-1} \sin k(R-b) \cosh K(R+b) \\
\chi'(k,R) & = & -k \sin k(R-b) K^{-1} \sinh K(R+b) \\
&&\hspace{.2in} +\cos k(R-b) \cosh K(R+b). 
\end{array}    \label{102}
\end{equation}
These expressions can be inserted in the equations for
$\rho,\widetilde{\rho},\tau,\widetilde{\tau}$, Eqs.~(\ref{53}) to
(\ref{54a}), to obtain the scattering amplitudes.  

Note that $K^2 = V_0 - k^2$ is a real symmetric matrix and may therefore
be diagonalized by an orthogonal transformation $U$, giving $K_D^2 =
UK^2U^T$.  The diagonal matrix $K_D$ has the square root of the
diagonal elements of $K_D^2$ along its diagonal.  Thus $K = U^TK_DU$,
which is a symmetric matrix.  Consequently, $\phi, \phi', \chi, \chi'$
are symmetric matrices, which leads to $\tau = \widetilde{\tau}$.
Furthermore, since each of the wave-function matrices or their
derivatives at $R$ is a power series (or polynomial) of the matrix
$K$, the wave-function matrices commute.  It is not difficult to show
that in the case of $b=R$, i.e., when the potential function has even
parity, $\rho(k) = \widetilde{\rho}(k)$.
           
The threshold behaviour of the transition amplitudes for a
potential lacking specific parity can be studied explicitly with this model.
Since Levinson's theorem involves the trace of the amplitudes at zero
energy, we need to consider the diagonalized forms of the amplitudes
only.  We diagonalize each of the wave-function matrices of Eq.~(\ref{102})
using the same orthogonal matrix $U$ for each, and we denote the 
diagonal matrices at zero energy as
\begin{equation}
\begin{array}{lll}
\phi_D(0,R)  & = & \mbox{diag}(p_1,\dots,p_N)  \\
\phi'_D(0,R) & = & \mbox{diag}(\lambda_1,\dots,\lambda_N)  \\
\chi_D(0,R)  & = & \mbox{diag}(x_1,\dots,x_N) \\
\chi'_D(0,R) & = & \mbox{diag}(x'_1,\dots,x'_N). \\
\end{array} \label{103}
\end{equation}
When there is no half-bound state, we obtain, by inserting these
expressions into Eqs.~(\ref{53}) and (\ref{54a}), 
\begin{equation}
\rho(0) = \rho_D(0) = -{\bf 1} \hspace{.2in} \mbox{and} \hspace{.2in}
\tau(0)=\tau_D(0) = {\bf 0}. \label{104}
\end{equation}
When there are $n$ half-bound states and $\det \phi'(0,R) = 0$,
we write $\phi'_D(0,R) =
\mbox{diag}(0,\dots,0,\lambda_{n+1},\dots,\lambda_N)$.  
Using Eq.~(\ref{11}) we find that 
\begin{equation}
{\phi^\dagger}'(k,x) \chi(k,x) - \phi^\dagger(k,x) \chi'(k,x) = -{\bf 1}.
\label{105} \\
\end{equation}
In general the matrices $\phi$ and $\chi$ are real and for the
constant potential matrix they are symmetric as well.  Thus
\begin{eqnarray}
 \phi'(0,R) \chi(0,R) - \phi(0,R) \chi'(0,R) & = & \nonumber \\
 \phi_D'(0,R) \chi_D(0,R) - \phi_D(0,R) \chi_D'(0,R) & = & -{\bf 1}. 
\label{106}
\end{eqnarray}
From this relation it follows that $x_i' = 1/p_i$ for $i=1,\dots,n$,
and furthermore
\begin{equation}
\rho_D(0) =  \mbox{diag}\left(\frac{1-p_1^2}{1+p_1^2}, \dots, 
\frac{1-p_n^2}{1+p_n^2}, -1, \dots, -1 \right) \label{107} 
\end{equation}
\begin{equation}
\widetilde{\rho}_D(0) = \mbox{diag}\left(-\frac{1-p_1^2}{1+p_1^2}, \dots, 
-\frac{1-p_n^2}{1+p_n^2}, -1, \dots, -1 \right) \label{107a}
\end{equation}
\begin{eqnarray}
\tau_D(0) & = & \widetilde{\tau}_D(0) \nonumber \\
& = & \mbox{diag}\left(\frac{2p_1}{1+p_1^2}, \dots,
\frac{2p_n}{1+p_n^2}, 0, \dots , 0\right). \label{108}
\end{eqnarray} 
Clearly the relation (\ref{83}) is satisfied by Eqs.~(\ref{107}) 
and (\ref{107a}).
For the parity invariant potential obtained by setting $b=R$ in
Eq.~(\ref{102}) ,
$\phi(0,R)=\chi'(0,R)$.
Hence $p_i=x_i'$ for $i=1,\dots,N$, and it follows that $p_i^2=1$.
Such a potential therefore yields transition amplitudes of the form
\begin{eqnarray}
\rho_D(0) & = & \widetilde{\rho}_D(0)=\mbox{diag}(0,\dots,0,-1,\dots,-1) 
\mbox{ and} \nonumber \\
\tau_D(0) & = & \mbox{diag}(\pm 1,\dots,\pm 1,0,\dots,0). \label{109}
\end{eqnarray}
When the $i^{\rm th}$ diagonal element of $\phi_D(0,x)$ 
is an even (odd) function, then the $i^{\rm th}$ diagonal element of
$\tau_D(0)$ will have a plus (minus) sign with the one.  The
converse is not necessarily true.

Consider the special case of $N=1$.  For the
parity invariant potential with a half-bound state, one has
$\rho(0)=0$ and $\tau(0)=\pm 1$.  The plus sign corresponds to
$\phi(0,x)$ being an even solution and the negative sign to it being
an odd solution.  When there is no half-bound state, then
$\rho(0)=-1$ and $\tau(0)=0$.  In the case of a potential without
definite parity, 
$\rho(0)=\widetilde{\rho}(0)=-1$ and $\tau(0)=\widetilde{\tau}(0)=0$
when the potential does not support a half-bound state.  When
there is a half-bound state, $\tau(0)$ and
$\widetilde{\tau}(0)$ are not equal to zero, nor are the $\rho$'s
equal to $-1$.  However, the sum of the $\rho$'s is an integer, i.e.,
$\rho(0)+\widetilde{\rho}(0) = 0$.  These results, which are clearly
valid for the square-well potential, are 
actually valid for any $N=1$ potential function.

\subsection{Models involving delta-function potentials}
\label{onedelta}

We now consider two examples involving delta functions for which
results can be obtained in closed form.
The results  exhibit qualitative
features which are also found in much more complicated examples.
First we examine the case of a single delta-function matrix potential
positioned at the origin.  Then we will use the factorization formula
derived earlier to look at the case for which there are two
delta-function matrices symmetrically positioned on both sides of the
origin.

\subsubsection{Delta function at the origin}

We write the Schr\"{o}dinger equation for this case as \equation
\left( -\frac{d^2}{dx^2} +\delta (x)\lambda\right) \psi = k^2\psi ,
\label{schroddel} \endequation where $\lambda$ is an $N\times N$
symmetric matrix, and $\psi$ can be taken to be either a column vector
solution or a solution matrix.  The former approach will be used when
we consider bound states and the latter when we examine scattering
solutions.

First consider the scattering solutions.  Since the potential has even
parity, we immediately have the result that $\widetilde{\rho} =\rho$
and $\widetilde{\tau}=\tau$.  Thus we need only consider the solution
with the incident wave from the left,
\begin{equation}
	\psi(k,x)=\cases{{\bf 1}e^{ikx}+\rho e^{-ikx},& $x\le 0$\cr
			\tau e^{ikx},& $x\ge 0$.\cr}
	\label{bcdel}
\end{equation}
Here we see the utility of working directly with a matrix of column
eigenvectors (as opposed to working with individual column vectors);
$\rho$ and $\tau$ may be solved for directly in terms of matrix 
operations.  The scattering amplitudes are
\begin{eqnarray}
	\rho(k) & = & (2ik-\lambda)^{-1}\lambda ,\\
	\tau(k) & = & {\bf 1}+\rho(k) =2ik(2ik-\lambda)^{-1} .
\end{eqnarray}
Of particular interest to us, due to its connection with the version
of Levinson's theorem given in Eq.~(\ref{48}), is the quantity 
$\mbox{Tr}[\rho(0)+\widetilde{\rho}(0)]$.  If $\lambda^{-1}$ exists, then
$\rho(0)=-\bf{1}$ and $\mbox{Tr}[\rho(0)+\widetilde{\rho}(0)]=
2\mbox{Tr}[\rho(0)]=-2N$, as expected.  If $\lambda$ is not invertible, 
however, we must be a bit more careful.

In order to determine the significance of the noninvertibility of
$\lambda$, consider the bound-state solutions of
Eq. (\ref{schroddel}).  Setting $\alpha^2 = -k^2$ and insisting that
$\alpha \ge 0$, we find that the column eigenvector for the bound
state is
\begin{equation}
        \Psi_b(\alpha,x)=\cases{A e^{\alpha x}, & $x\le 0$\cr
                        A e^{-\alpha x}, & $x\ge 0$,\cr}
\end{equation}
where $A$ is a normalized column matrix.  By integrating
Eq.~(\ref{schroddel}) over an infinitesimal interval including the
origin, we obtain an expression between the derivatives of
$\Psi_b(\alpha,x)$ on both sides of the origin, which leads to the
relation 
\equation
	(2\alpha +\lambda)A=0.
\endequation
In order to avoid the trivial solution, we demand that
\equation
	\det(2\alpha +\lambda) = 0.
	\label{detlam}
\endequation
The non-negative values of $\alpha$ which solve the above equation
define the bound-state energies.  Clearly there is at least
one half-bound state if $\det\lambda=0$.

Returning to the scattering problem, we find that the easiest way
to proceed is to first diagonalize the matrix $\lambda$.
Since $\lambda$ is real and symmetric, the diagonalization can
be accomplished by using an orthogonal matrix $U$, so that
\equation
	\lambda_D = U\lambda U^{-1},
\endequation
where $\lambda_D$ is diagonal and $U^{\rm{T}}=U^{-1}$.
If we now define 
\equation 
        \psi_D \equiv U\psi U^{-1}, 
\endequation 
we see that Eq. (\ref{schroddel}) may be rewritten as
\equation
        \left( -\frac{d^2}{dx^2} +\delta (x)\lambda_D\right)
                \psi_D= k^2\psi_D .
        \label{schroddeld}
\endequation
The orthogonal transformation similarly transforms the boundary
conditions, Eq. (\ref{bcdel}), to 
\begin{equation}
        \psi_D(k,x)=\cases{{\bf 1}e^{ikx}+\rho_D(k) e^{-ikx}, & $x\le 0$\cr
                        \tau_D(k) e^{ikx}, & $x\ge 0$.\cr}
        \label{bcdeld} 
\end{equation}
We see here another advantage of working directly with square matrices.
If we had been working with column vector wave functions, the transformed
wave functions would have been given by $U\Psi$, so that the 
normalization of the incoming wave would in general have been changed.
Working with $N\times N$ wave-function matrices gives the above result that
the form of the boundary conditions is unchanged under the transformation.
In fact the transformed wave function is itself a diagonal matrix, and
we essentially have $N$ decoupled copies of the problem with no coupling,
with (possibly) different potential strengths.\footnote{Note 
that this same trick can be employed any time the potential is
of the form $V(x)=v(x)M$,
where $M$ is a real symmetric matrix.  Diagonalizing $M$ gives $N$
decoupled systems with potentials $V_i(x)=m_i v(x)$, $i=1,\ldots, N$, 
where the $m_i$ are the eigenvalues of the matrix $M$.}

Suppose now that $\det\lambda=0$.  Then it follows that $\lambda$ has
at least one zero eigenvalue.  Let us again suppose that there are 
in fact $n$ zero eigenvalues, so that 
\equation
	\lambda_D= {\rm diag}(0,\ldots ,0,\lambda_{n+1},\ldots,\lambda_N),
\endequation
where the $\lambda_i$, $i=n+1,\ldots,N$, are the remaining (nonzero) 
eigenvalues.  Then the diagonalized reflection and transmission 
amplitude matrices are given by
\equation
        \rho_D(k)= {\rm diag}\left(0,\ldots ,0,\frac{\lambda_{n+1}}
          {2ik-\lambda_{n+1}},\ldots,\frac{\lambda_N}{2ik-\lambda_N}\right)
\endequation 
and
\equation 
        \tau_D(k)= {\rm diag}\left(1,\ldots ,1,\frac{2ik} 
                {2ik-\lambda_{n+1}},\ldots,\frac{2ik}{2ik-\lambda_N}\right),
\label{426}
\endequation  
so that 
\equation
	{\rm Tr}[\rho(0)+{\widetilde \rho}(0)]=
		2{\rm Tr}[\rho_D(0)]=-2(N-n) .
\endequation 
Thus we see in this example how the trace of 
$\rho(0)+{\widetilde \rho}(0)$ keeps track of the number of 
half-bound states in the system.  In fact it is easy to
verify that Levinson's theorem holds for the coupled system,
since it holds separately for each decoupled equation of the diagonalized 
problem.\footnote{The proof follows on noting that the determinant of the $S$
matrix is unchanged under the transformation which diagonalizes
$\rho$ and $\tau$.}

It is instructive to consider the relation
\begin{equation}
2ik\tau^{-1}_D(k)= \mbox{diag}(2ik,\dots,2ik,2ik-\lambda_{n+1},\dots,
2ik-\lambda_N),
\end{equation}
which follows from Eq.~(\ref{426}).  It demonstrates for this model
that in the limit as $k\rightarrow 0$ the expression 
Eq.~(\ref{68a}) is real, as expected.

\subsubsection{Potential with two delta functions}

We now turn to a slightly more complicated example, in which there
are two delta-function matrix potentials, one at $x=a$ and the other
at $x=-a$.  The $N=1$ version of this model was studied by Senn\cite{Senn88}.
The Schr\"{o}dinger equation for this case is given by
\equation
        \left( -\frac{d^2}{dx^2} +\delta (x+a)\lambda
		+\delta(x-a){\widetilde \lambda}\right)
                \psi = k^2\psi ,
        \label{schroddeldel}
\endequation
with boundary conditions
\begin{equation}
        \psi(k,x)=\cases{{\bf 1}e^{ikx}+\rho(k) e^{-ikx}, & $x\le -a$\cr
                        \tau(k) e^{ikx}, & $x\ge a$,\cr}
\end{equation}
for the wave incident from the left and
\begin{equation} 
        {\widetilde \psi}(k,x)=\cases{{\widetilde \tau}(k) e^{-ikx},& 
                          $x\le -a$ \cr
{\bf 1}e^{-ikx}+{\widetilde \rho}(k) e^{ikx}, & $x\ge a$,\cr}
\end{equation}
for the wave incident from the right.
Rather than solve the Schr\"{o}dinger equation again, we may now
simply substitute the results of the previous section into the 
factorization formula, Eq.~(\ref{facteq}).\footnote{Some care must be
taken with the reflection amplitude matrices, for they acquire
a phase when the potential is translated.  The transmission amplitude
matrices are, however, unchanged.}  An evaluation of the resulting
expressions yields 
\begin{eqnarray}
	\rho(k) & = & \left(\lambda e^{-4ika}+(2ik+\lambda)
		  (2ik-{\widetilde \lambda})^{-1}{\widetilde \lambda}
	          \right) \nonumber \\
&& \hspace{.2in} \Gamma^{-1}(k,a;\lambda,{\widetilde \lambda})
		  \lambda^{-1},\\
	\tau(k) & = & -4k^2e^{-2ika}(2ik-{\widetilde \lambda})^{-1}
		  \Gamma^{-1}(k,a;\lambda,{\widetilde \lambda})
		  \lambda^{-1} ,\\
	{\widetilde \rho(k)} & = & (2ik-{\widetilde \lambda})^{-1}
		  \Gamma^{-1}(k,a;\lambda,{\widetilde \lambda})
\nonumber \\
&& \hspace{.2in}  \left((2ik-\lambda)\lambda^{-1}{\widetilde \lambda}
		  e^{-4ika}+2ik+{\widetilde \lambda}\right) ,\\
	{\widetilde \tau(k)} & = & \tau^{T}(k) ,
\end{eqnarray}
where
\equation
	\Gamma(k,a;\lambda,{\widetilde \lambda}) =
		(2ik-\lambda)\lambda^{-1}e^{-2ika}-
		(2ik-{\widetilde \lambda})^{-1}{\widetilde \lambda}e^{2ika} .
	\label{gamdef}
\endequation
Let us assume for the moment that both $\lambda^{-1}$ and ${\widetilde
\lambda}^{-1}$ exist so that the above expressions are well defined.
(For $k>0$, it is actually sufficient that only one or the other
exists --
it is possible to rewrite the expressions so that they contain only
${\widetilde \lambda}^{-1}$ and not $\lambda^{-1}$.)
Performing a Taylor expansion of $\rho$ and ${\widetilde \rho}$ for small 
$k$, we find that in the typical case 
$\rho(0)={\widetilde \rho}(0)=-{\bf 1}$, so that
$\mbox{Tr}[\rho(0)+{\widetilde \rho}(0)]=-2N$, as expected.  The atypical
case is defined by the condition $\det(\lambda^{-1}+{\widetilde \lambda}^{-1}
+2a)=0$, which, as we shall see, is also the condition for a 
half-bound state.

Let us then work out the bound-state condition.  This may be done in 
a manner similar to that for the single delta-function case to obtain
\equation
	\det\Gamma(i\alpha,a;\lambda,{\widetilde \lambda}) = 0.
	\label{detlam1}
\endequation
Solutions of Eq.~(\ref{detlam1}) with $\alpha>0$ correspond to 
bound states.  
As $\alpha
\rightarrow 0$, Eq.~(\ref{detlam1})
yields
the half-bound-state condition,
\equation 
        \det(\lambda^{-1}+{\widetilde \lambda}^{-1} 
            +2a)=0 .
	\label{zebll}
\endequation
Alternatively, if we employ the wave functions $\phi$ and $\chi$ of 
Sec.~\ref{threshold}
for the model potential and use Eq.~(\ref{60}) as the condition
for the bound state, we obtain the equation
\begin{equation} 
\det\left(\left[(\lambda + 2\alpha)(\widetilde{\lambda} + 2\alpha) -
\lambda\widetilde{\lambda} e^{-4\alpha a}\right]/2\alpha\right) = 0,
\end{equation}
which in the limit as $\alpha$ approaches zero reduces to
\begin{equation} 
\det(\lambda + \widetilde{\lambda} + 2a\lambda\widetilde{\lambda})=0.
\end{equation}
This equation is preferred over Eq.~(\ref{zebll}) since it is not 
artificially singular when
one of the inverse matrices does not exist.

\begin{figure}[htb]
\begin{center}
\leavevmode
\epsfxsize3.375in
\centerline{\epsffile{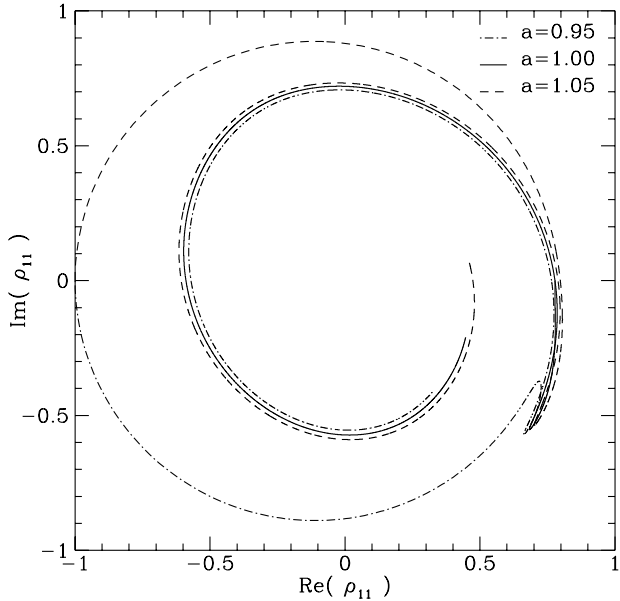}}
\end{center}
\caption{Plot of the Im($\rho_{11}(k)$) versus the Re($\rho_{11}(k)$)
for $k=0$ to $k=5$ for double delta-function matrix potential.  For the
$a=1.05$ case the curve reverses the direction of travel around the
origin when $k\simeq 1.1$.}
\label{fig1}
\end{figure}

Let us now consider an explicit example with $N=2$.  Since one of the
two matrices $\lambda$ or ${\widetilde \lambda}$ may always be
diagonalized by an orthogonal transformation, we will let $\lambda$ be
diagonal right from the start.  An example which gives a half-bound
state for $a=1$ is one for which 
\equation 
\lambda =
\left(\begin{array}{cc} -\frac{1}{2} & 0 \\ 0 & -1 \end{array}\right)
,\hspace{.5in} {\widetilde \lambda} = \left(\begin{array}{cc} -6 & -2
\\ -2 & -1 \end{array}\right) .  
\endequation 
Fig.~\ref{fig1} shows a
parametric plot of $\rho_{11}(k)$ as a function of $k$ in the complex
plane for the cases $a=0.95$, $a=1.00$, and $a=1.05$.  In the two
typical cases ($a=0.95, 1.05$), $\rho_{11}(0)=-1$, while for the
atypical case ($a=1$), $\rho_{11}(0)=0.777\ldots$.  This is then the
analog of Senn's ``threshold anomaly'' for the generalized matrix
version of his model\cite{Senn88}.  Examination of the other diagonal
reflection amplitudes yields the expected result that
$\mbox{Tr}[\rho(0)+{\widetilde \rho}(0)]$ is equal to $-4$ in the
typical case and $-2$ in the case where one half-bound state exists.

The behaviour of $\rho_{11}(k)$ as a function of $k$ may strike the
reader as being somewhat peculiar: for sufficiently large $k$ as $k$
increases, $\rho_{11}(k)$ traces out a never-ending counter-clockwise
spiral towards the origin.  A plot of the argument of $\rho_{11}(0)$
for the three cases would show that the phase shifts are not bounded
-- they keep on increasing to infinity.  This peculiar feature does
not exist when there is no coupling (the phase shifts
are bounded due to Levinson's theorem), but is a rather generic
feature of $N>1$ models.  The important thing to bear in mind when
$N>1$ is that the phase shift which obeys Levinson's theorem is
defined as being proportional to the logarithm of the determinant of
the $S$ matrix.  This phase shift can in general be a nontrivial
function of the ``physical'' phase shifts associated with the
scattering amplitudes.

\begin{figure}[htb]
\begin{center}
\leavevmode
\epsfxsize3.375in
\centerline{\epsffile{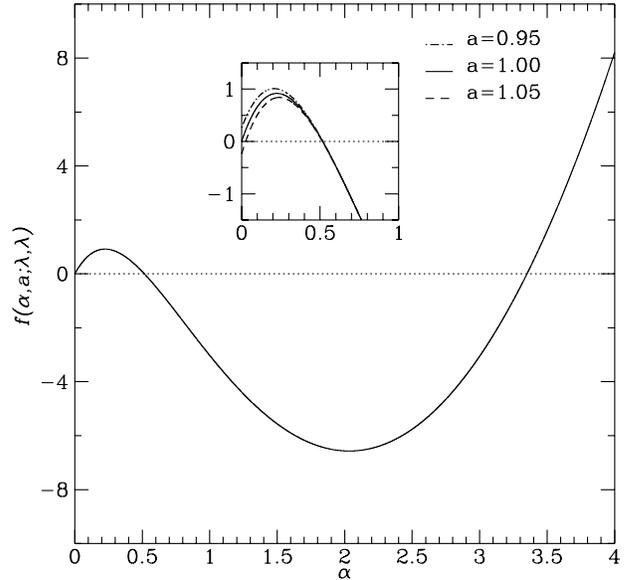}}
\end{center}
\caption{The determinant function for the double delta-function
potential graphed as a function of $\alpha$.}
\label{fig2}
\end{figure}

Fig.~\ref{fig2} shows a plot of
\equation
	f(\alpha,a;\lambda,{\widetilde \lambda})
		=\det\left( [(\lambda+2\alpha)(\widetilde{\lambda} +
2\alpha) - \lambda\widetilde{\lambda} e^{-4\alpha a}]/2\alpha\right),
\endequation
as a function of $\alpha$ for $a\approx 1$. 
 The inset in this figure shows an expanded view
of the function near the origin for the three cases $a=0.95, 1.00$,
and $1.05$.  In addition to the two regular bound states that all
three cases possess (near $\alpha=0.5164$ and $\alpha=3.3508$),
the $a=1.05$ case has an extra bound state near $\alpha=0.0259$,
and the  $a=1.00$ case has a new bound state just emerging
at $\alpha=0$.

Finally, Fig. 3 shows a plot of the ``Levinson's theorem'' phase shift
as a function of $k$ for the three cases.  Clearly this phase shift is
well-behaved and is bounded.  As the potential ``strength'' is
adjusted so that the system goes through a half-bound state the phase
shift at $k=0$ jumps by $\pi$ in two increments of $\pi/2$.

\begin{figure}[htb]
\begin{center}
\leavevmode
\epsfxsize3.375in
\centerline{\epsffile{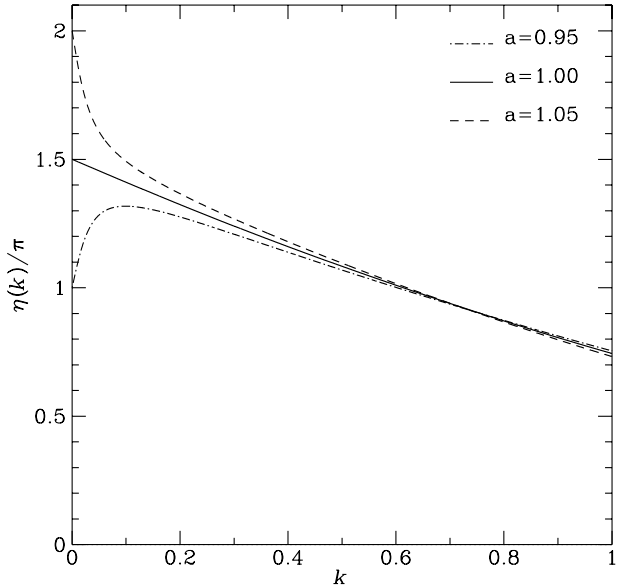}}
\end{center}
\caption{The phase of the $S$ matrix (divided by $\pi$) of the double
delta-function potential as a function of $k$.  For $k$ larger than
shown on the graph the three curves remain close to one another and
approach zero as $k\rightarrow\infty$.}
\label{fig3}
\end{figure}

\section{Discussion}
\label{sec:disc}

In this section we make a few observations.  The problem of
one-dimensional coupled-equation scattering using a representation of wave
functions which have incoming waves from the left or the right is
readily solvable.  Despite the advantages of a ``partial wave''
representation~\cite{Nogami95,Eberly65} or a parity-eigenstate
representation~\cite{DeBianchi94,VanDijk92} for parity-invariant
potential functions, our analysis (which is valid for any potential)
is quite manageable.

The use of wave function matrices (see Refs.~\cite{Newton55,Newton60}
for three-dimensional scattering and also \cite{VanDijk79,VanDijk79a}
for one-dimensional scattering) rather than column-vector wave
functions, leads to simplified notation for a number of relations,
e.g., the closure relation, Eq.~(\ref{35}).  One also finds that
performing a unitary transformation on the scattering wave function
matrix does not alter the normalization of the incoming waves, whereas
it does for column wave functions.  The introduction of the real
matrix wave function solutions $\phi$ and $\chi$ has two distinct
advantages.  In the first place the Schr\"{o}dinger equation for the
scattering problem can be reduced to a system of first-order
differential equations with one-point boundary conditions,
Eq.~(\ref{51b}).  The scattering amplitudes (and the $S$ matrix) are
algebraic expressions of these functions evaluated at $R$.
Furthermore, unlike the solution matrices $\psi$ and
$\widetilde{\psi}$, matrices $\phi$ and $\chi$ have linearly
independent columns at threshold and consequently are convenient for
investigating threshold behaviour.

Our starting point with the wave functions $\psi$ and
$\widetilde{\psi}$, which gives the definition of the reflection
amplitudes $\rho$ and $\widetilde{\rho}$, yields a generalized and
simplified understanding of threshold behaviour.  Whereas previous
work~\cite{Senn88,Nogami95} indicates that for parity-noninvariant
potentials the reflection amplitude at threshold can have noninteger
values, unlike that for parity-invariant potentials, we find that
$\mbox{Tr}[\rho(k) + \widetilde{\rho}(k)]$ at threshold is always an
integer (see Eq.~(\ref{83})).  The results of de
Bianchi~\cite{DeBianchi94}, however, already imply  such a relation, as well
as noninteger reflection amplitudes at threshold, for
some parity-noninvariant potentials with no coupling.

Finally, the phases of the reflection and transmission amplitudes of
the coupled system are not simple functions of $k$, as is the case for
uncoupled scattering, for which these phases satisfy the appropriate
form of Levinson's theorem.  The phase $\eta$ of the $S$ matrix, which
appears in Levinson's theorem, is in general a nontrivial function of
the phases of the scattering amplitudes.  Furthermore, the function
$\eta(k)$ is bounded, unlike the phases of the scattering amplitudes
which are not necessarily bounded.

\section*{Acknowledgments}
It is a pleasure to thank Professor Yuki Nogami for helpful
discussions.  KAK also benefited from discussions
with B. Leduc, J. Oppenheim and L. Paniak. 
The authors are grateful for support received from the
Natural Sciences and Engineering Research Council of Canada in the
form of Grant No. OGP0008672 (WvD) and an Undergraduate Research
Award (KAK). 

\appendix
\section{Some properties of solutions of the Schr\"{o}dinger equation}
In this appendix we determine the condition for linear independence of
the solutions of the Schr\"{o}dinger equation.  Consider the matrix
Schr\"{o}dinger equation
\begin{equation}
-\frac{d^2\psi}{dx^2} + V(x) \psi = k^2 \psi,  \label{A1}
\end{equation}
where $\psi$ is the $N\times N$ solution matrix whose columns are solutions to
Eq.~(2.1).  Suppose we have two
such solution matrices, $f$ and $\tilde{f}$.  We define a $2N\times
2N$ matrix functional
\begin{equation}
W(f,\tilde{f})=\left(\begin{array}{cc} ~f~ & ~\tilde{f}~ \\
                                       ~f'~ & ~\tilde{f}'~ 
  \end{array} \right), \label{A2}
\end{equation}
in which the prime indicates the derivative with respect to $x$.
Since $f$ and $\tilde{f}$ satisfy Eq.~(\ref{A1}) the 
matrix $W$ is a solution of the matrix equation,
\begin{equation}
W' = FW, \label{A3}
\end{equation}
where 
\begin{equation}
F = \left( \begin{array}{cc} {\bf 0} & ~{\bf 1}~ \\
                            V-k^2 &  {\bf 0} 
\end{array} \right).   \label{A4}
\end{equation}

\newtheorem{lemma}{Lemma}
\begin{lemma}
$(\det W)' = 0$ for all $x \in (-\infty,\infty)$.
\end{lemma}
{\em Proof:}  Let us write $f$ and
$\tilde{f}$ in terms of their $N$-component row vectors:
\begin{equation}
f=\left(\begin{array}{c} f_1 \\ f_2 \\ \vdots \\ f_N \end{array}
\right) \hspace{.3in}\mbox{ and } \tilde{f}=\left( \begin{array}{c}
\tilde{f}_1 \\ \tilde{f}_2 \\ \vdots \\ \tilde{f}_N \end{array}
\right). \label{A5}
\end{equation}
We then obtain~\cite[page 96]{Henrici77}
\begin{eqnarray}
 (\det W)' & = & \det\left(\begin{array}{cc} f'_1 & \tilde{f}'_1 \\
                                         f_2  & \tilde{f}_2  \\
                                      \vdots  & \vdots       \\
                                         f_N  & \tilde{f}_N  \\ 
                                         f'_1 & \tilde{f}'_1 \\
                                         f'_2 & \tilde{f}'_2 \\
                                      \vdots  & \vdots       \\
                                         f'_N & \tilde{f}'_N 
\end{array} \right) + \cdots + 
             \det\left(\begin{array}{cc} f_1 & \tilde{f}_1 \\
                                         f_2  & \tilde{f}_2  \\
                                      \vdots  & \vdots       \\
                                         f'_N  & \tilde{f}'_N  \\ 
                                         f'_1 & \tilde{f}'_1 \\
                                         f'_2 & \tilde{f}'_2 \\
                                      \vdots  & \vdots       \\
                                         f'_N & \tilde{f}'_N 
\end{array} \right) \nonumber \\
& + & \det\left(\begin{array}{cc} f_1 & \tilde{f}_1 \\
                                         f_2  & \tilde{f}_2  \\
                                      \vdots  & \vdots       \\
                                         f_N  & \tilde{f}_N  \\ 
                                         f''_1 & \tilde{f}''_1 \\
                                         f'_2 & \tilde{f}'_2 \\
                                      \vdots  & \vdots       \\
                                         f'_N & \tilde{f}'_N 
\end{array} \right) + \cdots  +
             \det\left(\begin{array}{cc} f_1 & \tilde{f}_1 \\
                                         f_2  & \tilde{f}_2  \\
                                      \vdots  & \vdots       \\
                                         f_N  & \tilde{f}_N  \\ 
                                         f'_1 & \tilde{f}'_1 \\
                                         f'_2 & \tilde{f}'_2 \\
                                      \vdots  & \vdots       \\
                                         f''_N & \tilde{f}''_N 
\end{array} \right). \nonumber \\
  \label{A6}
\end{eqnarray}
The first $N$ determinants in the sum on the right are zero because
they have two equivalent rows.  In order to show that the remaining
terms are also zero, we write the Schr\"{o}dinger equation,
Eq.~(\ref{A1}), as
\begin{equation}
\psi''(x) = M(x) \psi(x), \label{A7}
\end{equation}
where $M(x)$ is an $N\times N$ matrix.  Furthermore we write $M(x)$ in
terms of row vectors:
\begin{equation}
M(x) = \left( \begin{array}{c} m_1(x) \\
                               m_2(x) \\
                               \vdots \\
                               m_N(x)
\end{array} \right).  \label{A8}
\end{equation} 
Then the matrix in the $i^{\mbox{th}}$ second-derivative term of
Eq.~(\ref{A6}) may be written as
\begin{equation} 
\left(\begin{array}{cc} f_1 & \tilde{f}_1  \\
                       \vdots & \vdots \\
                        f_N & \tilde{f}_N \\
                        f'_1 & \tilde{f}'_1 \\
                        \vdots & \vdots \\
                        f''_i & \tilde{f}''_i \\
                        \vdots & \vdots \\
                        f'_N & \tilde{f}_N 
\end{array} \right) = 
\def\temp{\multicolumn{1}{c|}{}}
\left( \begin{array}{ccc} & \temp & \\
                          & \temp & \\
                   {\bf 1} & \temp & {\bf 0} \\
                          & \temp &  \\
                          & \temp &  \\ \cline{1-3}
                   m_i(x) & \temp & \begin{array}{ccccccc}
1 & & & & & & \\
  & \ddots & & & & & \\
  & & 1 & & & & \\
  & & & 0 & & & \\
  & & & & 1 & & \\
  & & & & & \ddots & \\
  & & & & & & 1
\end{array}
\end{array} \right) \left( W \right). \label{A9}
\end{equation}
The determinant of the matrix on the right side of Eq.~(\ref{A9}) is zero, and consequently the $(\det
W)' = 0$.   The $\det W $ is a constant function of $x$.

\begin{lemma}
The solutions contained in the columns of $f$ and
$\tilde{f}$ are linearly independent if and only if $\det W \neq 0$
for all $x\in (-\infty,\infty)$.
\end{lemma}
{\em Proof:}  First suppose that $\det W \neq 0$.  We consider a
linear combination of solutions which is equal to the trivial
solution,
\begin{equation}
f {\bf h} + \tilde{f} \tilde{\bf h} = {\bf o} \hspace{.2in} \mbox{ for
all }x,
\end{equation}
where ${\bf h}$ and $\tilde{\bf h}$ are $N$-component column vectors
of constants and ${\bf o}$ is the $N$-component zero column vector.  A
similar relation holds for the derivatives of $f$ and $\tilde{f}$, so
that
\begin{equation}
W{\bf c} = {\bf o},
\end{equation}
where 
\begin{equation} 
{\bf c} = \left( \begin{array}{c} {\bf h} \\ \tilde{\bf h} \end{array}
\right), 
\end{equation} 
and ${\bf c}$ and ${\bf o}$ are now $2N$-component column vectors. If
$\det W \neq 0$ for some $x$, which according to the previous result
means it is nonzero for all $x$, ${\bf c} = {\bf o}$ and the column
solutions contained in $f$ and $\tilde{f}$ are linearly independent.  

Suppose now that $\det W = 0$.  If $\det W = 0 $ for some $x = x_0$,
then the system of linear equations $W(x_0){\bf c}={\bf o}$ has a
nontrivial solution {\bf c}.  We form a column solution of the system
of differential equations, Eq.~(\ref{A3}), ${\bf w}(x) = W(x){\bf c}$
which vanishes at $x_0$.  This is the trivial solution of
Eq.~(\ref{A3}); {\bf w}(x) = {\bf o} for all $x$.  It follows that the
column solutions contained in $f$ and $\tilde{f}$ are linearly
dependent.

\section{Analytic properties of the solution matrices $\phi$ and
$\chi$}

We consider solution matrices $\phi(k,x)$ and $\chi(k,x)$ of
Eq.~(\ref{A1}) with boundary conditions $\phi(k,-R)=\chi'(k,-R)={\bf
1}$ and $\phi'(k,-R)=\chi(k,-R)={\bf 0}$, where $R$ is the range of
the potential.  According to a theorem of Poincar\'{e} an ordinary
differential equation containing an entire function of some parameter
has solutions which are entire functions of the parameter provided
these solutions have boundary conditions which are independent of the
parameter.  We will show that $\phi$ and $\chi$ are entire functions
of $k$, following a similar derivation for partial-wave solutions in
three-dimensional scattering~\cite{Newton60,Goldberger64}.

It is straightforward to verify that the matrix functions $\phi$ and
$\chi$ with the given boundary conditions are solutions of 
integral equations of the Volterra type,
\begin{eqnarray} 
\phi(k,x) & = & {\bf 1} \cos k(x+R) \nonumber \\
 && \hspace{.2in} + \int_{-R}^x dx' \; \frac{\sin
k(x-x')}{k} V(x')\phi(k,x')   \label{A201} \\ 
\chi(k,x) & = & {\bf 1} \frac{\sin k(x+R)}{k} \nonumber \\
&& \hspace{.2in} + \int_{-R}^x dx' \; 
\frac{\sin k(x-x')}{k} V(x') \chi(k,x').  \label{A202}
\end{eqnarray} 
In order to show that each element of the solution matrices is an entire
function of $k$, we rewrite Eq.~(\ref{A201}) in the form
\begin{eqnarray}
\phi(k,x) & = & {\bf 1} \cos k(x+R) \nonumber \\
&& \hspace{.1in} + \int_{-R}^x dx' \int_0^{x-x'} dt
\; \cos kt \; V(x')\phi(k,x')   \label{A201A} 
\end{eqnarray}
We solve Eq.~(\ref{A201A}) by successive approximations of the form
\begin{equation}
\phi(k,x) = \sum_{s=0}^\infty \phi^{(s)}(k,x),  \label{A203}
\end{equation}
where
\begin{eqnarray}
&& \phi^{(s)}(k,x)  =  \int_{-R}^x dx' \int_0^{x-x'} dt \; \cos kt \;
V(x')\phi^{(s-1)}(k,x'),  \nonumber \\
&&\hspace{.2in} \mbox{for }  s\geq 1 \mbox{ and } \phi^{(0)}(k,x)  
=  {\bf 1}\cos k(x+R).
\end{eqnarray}
Thus
\begin{eqnarray}
|\phi_{ij}^{(s)}(k,x)| & \leq &\int_{-R}^x dx' \int_0^{x-x'} dt \;
|\cos kt| \nonumber \\
&& \hspace{.1in} \displaystyle{\sum_l}|V_{il}(x')| 
|\phi_{lj}^{(s-1)}(k,x')|, \hspace{.1in} s\geq 1 
\end{eqnarray}
and
\begin{equation}
|\phi_{ij}^{(0)}(k,x)|=\delta_{ij}|\cos k(x+R)|.
\end{equation}
Denoting $\Im{k}$ for the imaginary part of $k$ and using the relation 
$|\cos kt|\leq \cosh{\Im{k} t}$ for $t$ real,
we obtain upon iteration
\begin{eqnarray}
&& |\phi_{ij}^{(s)}(k,x)|   \leq   \sum_{i'j'}|\phi_{i'j'}^{(s)}(k,x)| 
\nonumber \\
& & \leq 
\left( \frac{\sinh(2\Im{k}R)}{\Im{k}}\right)^s \cosh(2\Im{k}R) \int_{-R}^x
dx_1 \int_{-R}^{x_1} dx_2 \; \cdots \nonumber \\
& & \int_{-R}^{x_{s-1}}dx_s \sum_{i',j',l_1,\dots,l_s} |V_{i'l_1}(x_1)||V_{l_1l_2}(x_2)|
\cdots|V_{l_sj'}(x_s)|. \nonumber \\
\end{eqnarray}
Since the integrand is a symmetric function under the interchange of
any pair $(x_i,x_j)$,
\begin{eqnarray}
&& |\phi_{ij}^{(s)}(k,x)|  \leq  
\left( \frac{\sinh(2\Im{k}R)}{\Im{k}}\right)^s \cosh(2\Im{k}R) \frac{1}{s!}
\nonumber \\
&& \sum_{i',j',l_1,\dots,l_s} \int_{-R}^x dx_1 \; |V_{i'l_1}(x_1)| \cdots 
 \int_{-R}^x dx_s \; |V_{l_sj'}(x_s)|.
\end{eqnarray}
Let 
\begin{equation}
M_0 = \max_{i,j} \int_{-R}^R dx' \; |V_{ij}(x')| < \infty.
\end{equation}
Then 
\begin{equation}
|\phi_{ij}^{(s)}(k,x)| \leq 
\left( \frac{\sinh(2\Im{k}R)}{\Im{k}}\right)^s \cosh(2\Im{k}R) 
\frac{N^{s+2}}{s!} M_0^s.
\end{equation}
Thus the series $\sum_s \phi_{ij}^{(s)}$ converges absolutely and
uniformly for $x \in [-R,R]$ and for every region in the complex $k$
plane.  To determine the existence of ${\displaystyle\frac{\partial \phi}{\partial
k}(k,x)}$, we differentiate Eq.~(\ref{A201A}) with respect to $k$,
\begin{eqnarray}
\frac{\partial \phi}{\partial k}(k,x) & = & -(x+R){\bf 1} \sin k(x+R)
\nonumber \\ 
& - & \int_R^x dx' \int_0^{x-x'} dt \; t\sin kt \; V(x') \phi(k,x')
\nonumber \\
 & & \hspace{-.3in} + \int_R^x
dx' \int_0^{x-x'} dt \; \cos kt \; V(x') \frac{\partial \phi}
{\partial k}(k,x'). 
\end{eqnarray}
When Eq.~(\ref{A203}) is differentiated with respect to $k$ it yields
\begin{equation}
\frac{\partial \phi}{\partial k}(k,x) = \sum_{s=0}^\infty
\frac{\partial \phi^{(s)}}{\partial k}(k,x), \label{A219}
\end{equation}
where now we have
\begin{equation}
\frac{\partial \phi^{(s)}}{\partial k}(k,x)  =  \int_R^x dx'
\int_0^{x-x'} dt \; \cos kt \; V(x') \frac{\partial \phi^{(s-1)}}{\partial
k}(k,x') 
\end{equation}
with
\begin{eqnarray}
\frac{\partial \phi^{(0)}}{\partial k}(k,x) & = & -(x+R) {\bf 1} \sin
k(x+R) \nonumber \\ 
& & \hspace{-.3in} - \int_R^x dx' \int_0^{x-x'} dt \; t \sin kt 
\; V(x') \phi(k,x').
\end{eqnarray}
It is not difficult to show that ${\displaystyle\left|\frac{\partial
\phi_{ij}^{(0)}}{\partial k}(k,x)\right|}$ is bounded, and the convergence of
series~(\ref{A219}) follows in the same manner as that of $\phi(k,x)$.
Since $\phi(k,x)$ and its derivative with respect to $k$ exist for all
$k$, $\phi(k,x)$ is an entire function of $k$.  Similarly $\chi(k,x)$
can be shown to be an entire function of $k$.

For real $k$ the behaviour of $\phi(k,x)$ and $\chi(k,x)$ as $k$
becomes very large can be determined by iterating Eqs.~(\ref{A201}) and
(\ref{A202}). Thus
\begin{eqnarray}
&& \phi(k,x) \;\;\;\; {\sim}\hspace{-13pt}\raisebox{-1.ex}{${\scriptstyle 
k\rightarrow\infty}$}\;\;
{\bf 1} \cos k(x+R) + \nonumber \\
&& \frac{1}{k}\int_{-R}^x dx' \sin
k(x-x') V(x') \cos k(x'+R) + O(1/k^2) \nonumber \\ \label{A210}
\end{eqnarray}
and
\begin{eqnarray}
&& \chi(k,x)\;\;\;\; {\sim}\hspace{-13pt}\raisebox{-1.ex}{${\scriptstyle 
k\rightarrow\infty}$}\;\;
{\bf 1} \frac{\sin k(x+R)}{k} + \nonumber \\
&& \frac{1}{k^2}\int_{-R}^x dx' \sin
k(x-x') V(x') \sin k(x'+R) + O(1/k^3). \nonumber \\  \label{A211}
\end{eqnarray}


\begin{thebibliography}{10}

\bibitem{Hauge89}
E.~H. Hauge and J.~A. St{\o}vneng, Reviews of Modern Physics {\bf 61},  917
  (1989).

\bibitem{DeBianchi94}
M.~S. {de Bianchi}, Journal of Mathematical Physics {\bf 35},  2719  (1994).

\bibitem{DeBianchi95}
M.~S. {de Bianchi} and M. {Di Ventra}, Journal of Mathematical Physics {\bf
  36},  1753  (1995).

\bibitem{Senn88}
P. Senn, American Journal of Physics {\bf 56},  916  (1988).

\bibitem{Barton85}
G. Barton, Journal of Physics: Mathematical and General {\bf 18},  479  (1985).

\bibitem{VanDijk92}
W. {van Dijk} and K.~A. Kiers, American Journal of Physics {\bf 60},  520
  (1992).

\bibitem{Nogami95}
Y. Nogami and C.~K. Ross, American Journal of Physics {\bf 64},  923  (1996).

\bibitem{Eberly65}
J.~H. Eberly, American Journal of Physics {\bf 33},  771  (1965).

\bibitem{Faddeev64}
L. Faddeev, American Mathematical Society Translations, Series 2 {\bf 65},  139
   (1964).

\bibitem{Newton80}
R.~G. Newton, Journal of Mathematical Physics {\bf 21},  493  (1980).

\bibitem{Newton83}
R.~G. Newton, Journal of Mathematical Physics {\bf 24},  2152  (1983).

\bibitem{Newton84}
R.~G. Newton, Journal of Mathematical Physics {\bf 25},  2991  (1984).

\bibitem{VanDijk79}
W. {van Dijk} and M. Razavy, International Journal of Quantum Chemistry {\bf
  16},  1249  (1979).

\bibitem{VanDijk79a}
W. {van Dijk} and M. Razavy, Canadian Journal of Physics {\bf 57},  1952
  (1979).

\bibitem{Lane58}
A.~M. Lane and R.~G. Thomas, Reviews of Modern Physics {\bf 30},  257  (1958).

\bibitem{Newton55}
R.~G. Newton and R. Jost, Il Nuovo Cimento {\bf 1},  590  (1955).

\bibitem{Mott65}
N.~F. Mott and H.~S.~W. Massey, {\em The theory of atomic collisions}, 3rd ed.
  (Oxford University Press, London, 1965).

\bibitem{Aktosun85}
T. Aktosun and R.~G. Newton, Inverse Problems {\bf 1},  291  (1985).

\bibitem{Newton77}
R.~G. Newton, Journal of Mathematical Physics {\bf 18},  1348  (1977).

\bibitem{Lancaster85}
P. Lancaster and M. Tismenetsky, {\em The Theory of Matrices with
  Applications}, 2nd  ed. (Academic Press, Orlanda, Florida, 1985).

\bibitem{Rellich69}
F. Rellich, {\em Perturbation Theory of Eigenvalue Problems} (Gordon and
  Breach, New York, 1969).

\bibitem{Aktosun92}
T. Aktosun, Journal of Mathematical Physics {\bf 33},  3865  (1992).

\bibitem{Rozman94a}
M. Rozman, P. Reineker, and R. Tehver, Physics Letters A {\bf 187},  127
  (1994).

\bibitem{Rozman94b}
M. Rozman, P. Reineker, and R. Tehver, Physical Review A {\bf 49},  3310
  (1994).

\bibitem{Sprung93}
D. Sprung, H. Wu, and J. Martorell, American Journal of Physics {\bf 61},  1118
   (1993).

\bibitem{Doyle75}
S.~D. Doyle, J.~S. Eck, W.~J. Thompson, and O.~L. Weaver, American Journal of
  Physics {\bf 43},  677  (1975).

\bibitem{Henrici77}
P. Henrici, {\em Applied and computational complex analysis} (John Wiley and
  Sons, New York, 1977), Vol.~2.

\bibitem{Newton60}
R.~G. Newton, Journal of Mathematical Physics {\bf 1},  319  (1960).

\bibitem{Goldberger64}
M.~L. Goldberger and K.~M. Watson, {\em Collision theory} (John Wiley and Sons,
  New York, 1964).

\end{thebibliography}
\end{document}